\documentclass[structabstract]{raa}
\usepackage{graphicx,times}
\usepackage{natbib,amssymb}
\usepackage{epsfig}
\usepackage{float}

\providecommand{\bm}[1]{\mbox{\boldmath$#1$\unboldmath}}

\begin{document}
   \title{The formation of neutron star systems through accretion-induced collapse in white-dwarf binaries}

   \volnopage{ {\bf 2020} Vol.\ {\bf XX} No. {\bf XXX}, 000--000}
   \setcounter{page}{1}

   \author{Bo Wang \inst{1,2,3,4} 
      \and 
      Dongdong Liu \inst{1,2,3,4}
          }
   \institute{Yunnan Observatories, Chinese Academy of Sciences,  Kunming 650216, China;
          {\it wangbo@ynao.ac.cn}\\
          \and
          Key Laboratory for the Structure and Evolution of Celestial Objects, Chinese Academy of Sciences, Kunming 650216, China
          \and
          Center for Astronomical Mega-Science, Chinese Academy of Sciences, Beijing 100012, China
          \and
          University of Chinese Academy of Sciences, Beijing 100049, China
              }              

   \date{Received; accepted}

\abstract {The accretion-induced collapse (AIC) scenario was proposed 40 years ago as an evolutionary end state of oxygen-neon white-dwarfs (ONe WDs), linking them to the formation of neutron star (NS) systems. However, there has been no direct detection of any AIC event so far, even though there exists a lot of indirect observational evidence. Meanwhile, the evolutionary pathways resulting in NS formation through AIC are still not well investigated. In this article, we review recent studies on the two classic progenitor models of AIC events, i.e., the single-degenerate model (including the ONe WD+MS/RG/He star channels and the CO WD+He star channel) and the double-degenerate model (including the double CO WD channel, the double ONe WD channel and the ONe WD+CO WD channel). Recent progress on these progenitor models is reviewed, including the evolutionary scenarios leading to AIC events, the initial parameter space for producing AIC events, and the related objects (e.g., the pre-AIC systems and the post-AIC systems). For the single-degenerate model, the pre-AIC systems (i.e., the progenitor systems of AIC events) could potentially be identified as supersoft X-ray sources, symbiotics and
cataclysmic variables (such as classical novae, recurrent novae, Ne novae and He novae) in the observations, whereas the post-AIC systems (i.e., NS systems) could potentially be identified as low-/intermediate-mass X-ray binaries, and the resulting low-/intermediate-mass binary pulsars, most notably millisecond pulsars. For the double-degenerate model, the pre-AIC systems are close double WDs with short orbital periods, whereas the post-AIC systems are single isolated NSs that may correspond to a specific kind of NSs with peculiar properties. We also review the predicted rates of AIC events, the mass distribution of NSs produced via AIC, and the gravitational wave (GW) signals from double WDs that are potential GW sources in the Galaxy in the context of future space-based GW detectors, such as LISA, TianQin and Taiji, etc. Recent theoretical and observational constraints on the detection of AIC events are summarized. In order to confirm the existence of the AIC process, and resolve this long-term issue presented by current stellar evolution theories, more numerical simulations and observational identifications are required.
 }
\titlerunning{The formation of NS systems through AIC in WD binaries}
\authorrunning{B. Wang \& D. Liu}
\maketitle


\section{Introduction}

Compact objects are the product of late stellar evolution, which 
can be divided into white-dwarfs (WDs), neutron stars (NSs) and black holes.
WDs are the endpoint of the evolution of low-/intermediate-mass stars, 
accounting for the final outcome of about 97\% of all stars in our Galaxy (e.g., Fontaine et al. 2001; Parsons et al. 2020). 
According to the underlying compositions, this kind of compact objects can be mainly
divided into He WDs, carbon-oxygen (CO) WDs and  oxygen-neon (ONe) WDs 
 (e.g., Paczy\'{n}ski 1970; Han et al. 1994; Eggleton 2006; Camisassa et al. 2019). Note that some transitional
hybrid WDs have been predicted by the stellar evolution theories but not confirmed in the observations. For example,  
He-rich WDs (i.e., HeCO WDs, in which a CO-rich core surrounded by a He-rich mantle; see Iben \& Tutukov 1985),  
CONe WDs (an unburnt CO core surrounded by a thick ONe zone; e.g., Denissenkov et al. 2013; Chen et al. 2014),
COSi WDs (an unburnt CO core surrounded by a Si-rich shell; see Wu et al. 2020), 
OSi WDs (see Wu \& Wang 2019; Wu et al. 2019) and Si WDs (see Schwab et al. 2016), etc.

It has been generally believed that CO WDs in binaries will form
type Ia supernovae (SNe Ia) when they grow in mass close to  the Chandrasekhar-mass limit 
(${M}_{\rm Ch}$; e.g., Hoyle \& Fowler 1960; Nomoto et al. 1984; 
Podsiadlowski 2010; Wang \& Han 2012).
However, ONe WDs are  predicted to collapse into NSs through 
electron-capture reactions  by Mg and Ne when they increase their mass close to ${M}_{\rm Ch}$,
in which the transformation from ONe WDs to NSs  is referred 
as the accretion-induced collapse (AIC) process that was proposed 40 years ago based on the stellar evolution theories
 (e.g.,  Nomoto et al. 1979; Miyaji et al. 1980; Taam \& van den Heuvel 1986; 
 Canal et al. 1990a; Nomoto \& Kondo 1991; Schwab et al. 2015, 2019). 
Some recent studies claimed that ONe WDs can 
experience explosive oxygen or neon burning  and form a subclass of SNe Ia finally (e.g., Marquardt et al. 2015; Jones et al. 2016; 
see also Isern et al. 1991). The final outcome of ONe WDs (collapse or explosion) is mainly determined by the competition between 
electron-capture reactions and nuclear burning. 
By simulating the long-term evolution of mass-accreting ONe WDs with different initial  masses  
and various mass-accretion rates, Wu  \& Wang (2018) found that  the central temperature of ONe WDs 
cannot reach the temperature for explosive oxygen or neon burning as the  neutrino-loss 
apparently increases with the central temperature. Thus,  the final outcome of ONe WDs with 
${M}_{\rm Ch}$ is to collapse into NSs.

NSs have been generally thought to be produced through core-collapse SNe and 
electron-capture SNe (e.g., van den Heuvel 2009). As a kind of electron-capture SNe, 
AIC events are expected to be  relatively faint optical transients  (e.g., Woosley \& Baron 1992). 
A small amount ($\sim10^{-3}-10^{-1}\,{M}_\odot$) of the ejecta mass  is 
predicted during the collapse, and  the $^{56}{\rm Ni}$ synthesized 
is posssibly $\sim10^{-4}-10^{-2}\,{M}_\odot$ (e.g., Fryer et al. 1999; Dessart et al. 2006, 2007; Metzger et al. 2009; Darbha et al. 2010).
It has been suggested that  AIC events  are fainter than that of a typical normal SNe Ia  by 5 mag or more,
and last for only a few days to a week (e.g., Piro \& Thompson 2014). 
Thus, they are relatively short-lived and most likely underluminous. This  indicates that this kind of objects are difficult to be discovered.
 Piro \& Kollmeier (2013) suggested that a transient radio source may appear after the AIC event, which can
last for a few months. 

Up to now, there has been no reported direct detection for such events, 
but there exists a lot of indirect evidence in the observations.
The remnants of AIC process are NS systems, which can be used to explain 
a variety of the observed troublesome NS systems that cannot be reproduced by core-collapse SNe
(e.g., Canal et al. 1990b;  Tauris et al. 2013; Liu \& Li 2017; Ablimit 2019), as follows:
\begin{enumerate}
\item [(1)] It has been suggested that some of the observed pulsars in globular clusters are significantly younger than the clusters themselves, 
which can be explained by the AIC scenario that could produce newborn NSs with small kicks (e.g., Boyles et al. 2011) . 
\item [(2)] The AIC scenario provides an alternative way to explain the formation of some millisecond pulsars (MSPs; 
e.g., Bhattacharya \& van den Heuvel 1991; Ivanova et al. 2008; Hurley et al. 2010; Freire \& Tauris 2014). 
Hurley et al. (2010) found that the rates of binary MSPs from the AIC scenario are comparable to those from core-collapse SNe. 
This indicates that the AIC scenario plays a key role for the production of binary MSPs. 
Note that Chen et al. (2011) argued that the AIC scenario cannot form eccentric binary 
MSPs with an orbital period of $\gtrsim$20\,d even though a high kick is considered.
This is because the formation of eccentric binary MSPs needs a high kick during the AIC, but 
the high kick is more likely to disrupt the binary when the orbit is too wide, 
forming more isolated MSPs finally (see Chen et al. 2011).
\item [(3)] The AIC scenario can be used to explain a large fraction of NSs in globular clusters and 
the formation of recycled pulsars with the observed low space velocities due to the small kicks 
and the small amount of mass-loss during the collapse (e.g., Bailyn \& Grindlay 1990; Kitaura et al. 2006; Dessart et al. 2006). 
\item [(4)] The recycling process of the AIC  scenario can be used to explain some mass-accreting NSs 
with strong magnetic fields in the low-mass X-ray binaries 
and some strong-magnetized pulsars with He WD companions, which  have undergone 
extensive mass-transfer  but have no too much matter accumulated  onto the surface of the NSs (e.g., 
Taam \& van den Heuvel 1986; van Paradijs et al. 1997;  Li \& Wang 1998; Xu \& Li 2009). 
\item [(5)] The AIC scenario has been invoked as an alternative  mechanism for the formation of low-/intermediate-mass X-ray binaries  
and low-/intermediate-mass binary pulsars (e.g., van den Heuvel 1984; 
Canal et al. 1990b; Nomoto \& Kondo 1991;  Li \& Wang 1998; Tauris et al. 2012;  Liu et al. 2018a). 
\end{enumerate}

Meanwhile, the AIC scenario has been proposed as a promising way to form some important events, as follows:
(1) The ejecta from the AIC process has been claimed as a possible source of r-process nucleosynthesis (e.g.,
Wheeler et al. 1998; Fryer et al. 1999; Qian \& Wasserburg 2007). 
(2) The AIC process has been proposed as a source of gravitational wave (GW) emission 
(e.g.,  Abdikamalov et al. 2010). 
(3) The AIC scenario may be a novel source of cosmic rays and of cosmological 
gamma-ray bursts (e.g., Usov 1992; Dar et al. 1992; Lyutikov \& Toonen 2017).
Piro \& Kollmeier (2016) recently argued that the AIC scenario can potentially form 
rapidly spinning  magnetars, in which the newborn magnetars
provide an attractive site for producing ultrahigh-energy cosmic rays.
(4) It has been suggested that 
AIC may provide an alternative way to result in fast radio bursts
(e.g., Moriya 2016; Cao et al. 2018; Margalit et al. 2019).
(5) Belczynski et al. (2018) recently argued that the AIC scenario can contribute to 
the formation of double NSs in the globular cluster dynamical channel. 
In this channel, a NS+ONe WD binary interacts with a CO WD, 
during which the ONe WD merges with the CO WD, leading to the formation of a NS via AIC
(see Fig. 3 of Belczynski et al. 2018).

Despite the potential importance of the AIC process,
it is still uncertain about their progenitor models. 
Meanwhile, the theoretical rates of AIC events remain highly uncertain.
Similar to the progenitor models of SNe Ia, two classic kinds of progenitor models for  AIC events have been proposed
so far (i.e., the single-degenerate model and the double-degenerate model), as follows:
\begin{enumerate}
\item [(1)]  \textit{The single-degenerate model}.
In this model, an ONe WD grows in mass by accreting matter from its 
non-degenerate companion. An AIC event may be produced
once the ONe WD increases its mass close to ${M}_{\rm Ch}$ 
(e.g., Nomoto \& Kondo 1991;  Yungelson \& Livio 1998; Ivanova \& Taam 2004; 
Tauris et al. 2013; Brooks et al. 2017; Liu et al. 2018; Wang 2018a; Ruiter et al. 2019). 
The single-degenerate model mainly produces NS binary systems through AIC.
\item [(2)]  \textit{The double-degenerate model}. 
This model is usually called merger-induced collapse, in which an AIC event results from the merging of double WDs 
with a total mass heavier than ${M}_{\rm Ch}$; the merging of double WDs was caused by the GW radiation that
results in its orbit to shrink
(e.g., Nomoto \& Iben 1985; Saio \& Nomoto 1985; Ruiter et al. 2019; Liu \& Wang 2020).
In addition to the orbital evolution caused by the GW radiation, 
the merger of double WDs in triple systems can be
caused by the Kozai-Lidov mechanism,
in which secular gravitational effects result in eccentricity oscillations in the inner binary
 (see Kozai 1962; Lidov 1962). For a recent review on the Kozai-Lidov mechanism, see Naoz (2016).
 The merger-induced collapse for double-degenerate binaries mainly produces single isolated NSs,
whereas the outcomes of triple systems via AIC likely form NS binaries with long orbital periods.
\end{enumerate}

In this article,  we mainly review 
recent progress on the currently most popular progenitor models  of AIC events, 
including the single-degenerate model in Sect.\,2 and the double-degenerate model in Sect.\,3. 
In Sect.\,4, we  review recent results of binary population synthesis for producing AIC events, including the
the predicted rates of AIC events in the Galaxy, the mass distribution of NSs via AIC, and the GW signals 
from double WDs that may produce AIC events.
We summarize recent  theoretical and observational  constraints on the detection of AIC events in Sect.\,5.
Finally, a summary is given in Sect.\,6.

\section{The single-degenerate model}

In this model, an ONe WD grows in mass by accreting H-/He-rich matter from its non-degenerate companion
that could be a main-sequence or a slightly evolved star (the ONe WD+MS channel), 
a red-giant star (the  ONe WD+RG channel), or even a He star (the  ONe WD+He star channel).
The accreted matter will be burned into C and O, and accumulate onto the surface of the ONe WD, 
resulting in the mass increase of the WD. 
When the ONe WD increases its mass close to ${M}_{\rm Ch}$, an AIC event may be produced.  
Note that the maximum stable mass for a WD is likely to be larger than ${M}_{\rm Ch}$ in consideration of rotation  
(e.g., Ostriker \& Bodenheimer 1968; Yoon \& Langer 2004; Wang et al. 2014).

Tauris et al. (2013) studied the binary computations of ONe WD+MS/RG/He star systems 
that may undergo the AIC process and then be recycled to produce binary pulsars.  
However, 
they only considered the case with the initial mass of the ONe WD
${M}^{\rm i}_{\rm ONe}=1.2\,M_{\odot}$. 
By adopting the optically thick wind assumption (see Hachisu et al. 1996), 
Wang (2018a)  investigated  the  ONe WD+MS/RG/He star channels for producing AIC events
 in a systematic way with different ${M}^{\rm i}_{\rm ONe}$. 
Compared with the results of Tauris et al. (2013),  the initial regions 
of AIC events obtained by Wang (2018a) are notably enlarged, 
mainly due to different prescriptions for the case of rapid mass-transfer process.
In addition, Wang et al. (2017) recently found that the CO WD+He star systems 
may also form NS systems through the AIC process 
when off-center carbon ignition happens on the surface of the CO WD, 
known as the CO WD+He star channel (see also Brooks et al. 2016).

\subsection{The ONe WD+MS channel}

\subsubsection{Evolutionary scenarios}

\begin{figure*}
\begin{center}
\epsfig{file=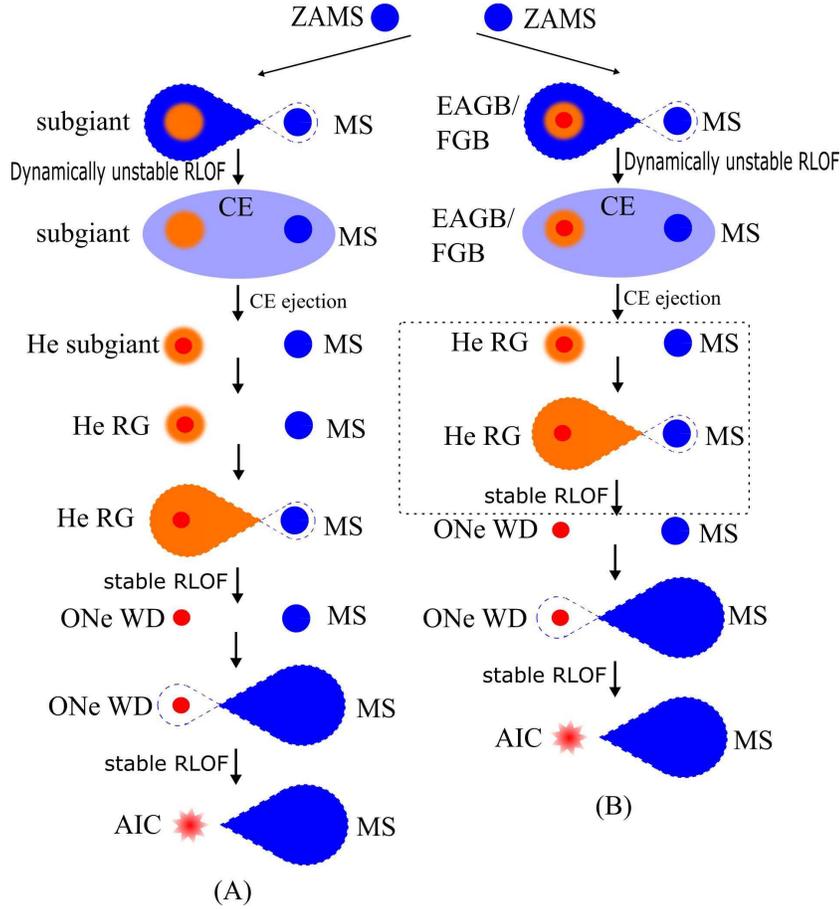,width=11cm} 
\caption{Evolutionary scenarios to ONe WD+MS systems that can form AIC events.
In Scenario B, some cases will not experience the second RLOF (dashed box).}
\end{center}
\end{figure*}

In this  channel, AIC events originate from the evolution of ONe WD+MS systems.
According to the evolutionary stage of the primordial primary (massive star) when it first fills its Roche-lobe, 
there are  two  evolutionary scenarios to form ONe WD+MS systems and then produce AIC events (see Fig.\,1), 
in which Scenario A is the main way to form AIC events  (see Wang 2018a).

\textbf{Scenario A}: The primordial primary fills its Roche-lobe at the Hertzsprung gap (HG) stage, 
leading to a case B mass-transfer process. In this case, a common-envelope (CE) may be formed because of 
the large mass-ratio or the convective envelope of the Roche-lobe filled star. 
After the ejection of CE, the primordial primary turns to be a He subgiant star. 
The primary continues to evolve and will form a He RG star when its central He is exhausted. 
The primary will expand quickly at this stage and will fill its Roche-lobe again, 
resulting in a stable mass-transfer process. 
After that, the primary turns to be an ONe WD,  forming an ONe WD$+$MS system. 
For this scenario, the initial parameters of the primordial binaries for producing AIC events are in the range of 
$M_{\rm 1,i}\sim8-11\, M_{\odot}$, 
$q=M_{\rm2,i}/M_{\rm1,i}\sim0.2-0.4$
and $P^{\rm i}\sim40-900\,\rm d$,  
in which $M_{\rm 1,i}$, $M_{\rm 2,i}$, $q$ and $P^{\rm i}$ are the initial masses of 
the primordial primary and the secondary, the initial mass-ratio  and the initial orbital period 
of the primordial systems, respectively.\footnote{The range of the initial parameters  were 
obtained under some specific assumptions for both the primordial binaries 
and binary interactions in the binary population synthesis studies (see Wang 2018a; Liu \& Wang 2020).  
Here, we set the CE ejection parameter $\alpha_{\rm CE}\lambda$ to be 1.5. 
If we adopt a low value of $\alpha_{\rm CE}\lambda$ (e.g., 0.5), 
the initial orbital period will become longer 
as the primordial binaries could release more orbital energy during the CE ejection so that  
they can evolve to  the initial parameter regions for producing AIC events.}
About three quarters of AIC events from the ONe WD$+$MS channel are produced through this scenario.

\textbf{Scenario B}: The primordial primary first fills its Roche-lobe at the early asymptotic 
giant branch (EAGB)  stage or the first giant branch (FGB) stage, 
and then a CE may be formed due to the dynamically unstable mass-transfer. 
If the CE can be ejected, the primary will turn to be a He RG star. 
The He RG continues to evolve and will quickly fill its Roche-lobe again, 
leading to a stable Roche-lobe overflow (RLOF) process. After the mass-transfer process, 
the primary becomes an ONe WD, resulting in the formation of an ONe WD$+$MS system. 
For this scenario, the initial parameters of the primordial binaries for producing AIC events are in the range of 
$M_{\rm 1,i}\sim6-8\, M_{\odot}$, 
$q\sim0.2-0.4$ 
and $P^{\rm i}\sim400-1000\,\rm d$.
About one quarter of AIC events from the ONe WD$+$MS channel are produced through this scenario.
Note that the mass-transfer process from a He RG to a MS shown in the dashed box of Fig.\,1 may be not indispensable in some cases.

\subsubsection{Parameter space for AIC events}

\begin{figure}
\begin{center}
\epsfig{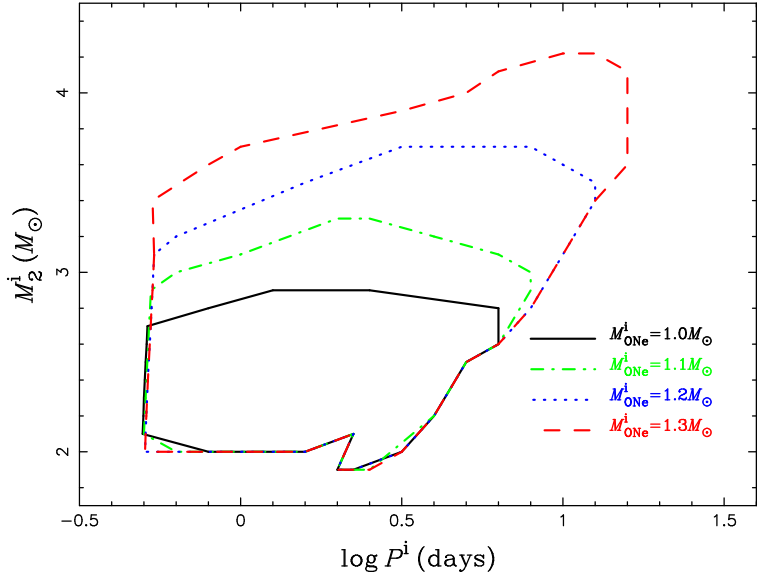} 
\caption{Initial parameter regions of ONe WD+MS systems for producing AIC events
in the $\log P^{\rm i}-M^{\rm i}_2$ plane with various ${M}^{\rm i}_{\rm ONe}$. 
Source: From Wang (2018a).}
\end{center}
\end{figure}

After the formation of ONe WD+MS systems, the ONe WD could accrete 
H-rich matter from its MS companion when it fills its Roche-lobe. 
The accreted H-rich matter is burned into He, and then the He is processed into C and O. 
The ONe WD may collapse into a NS when it grows in mass close to ${M}_{\rm Ch}$.
Fig.\,2 shows the initial parameter regions of AIC events for the ONe WD+MS channel in 
the $\log P^{\rm i}-M^{\rm i}_2$ plane with different ${M}^{\rm i}_{\rm ONe}$, 
where $M^{\rm i}_2$ is the initial mass of the MS star and $P^{\rm i}$ is the initial orbital period  of the  ONe WD+MS system. 
In order to produce AIC events, the ONe WD should have MS companions with initial masses of $\sim$2$-$4.2\,$M_{\odot}$ 
and initial orbital periods of $\sim$0.5$-$16\,d.

The boundaries of the initial regions in Fig.\,2 are mainly constrained by the following conditions:
(1)
The lower boundaries 
are set by a low mass-transfer rate that prevents the ONe WDs growing in mass to ${M}_{\rm Ch}$ during H-shell flashes. 
(2)
The upper boundaries are constrained by a high
mass-transfer rate owing to a large mass-ratio between the MS star
and the ONe WD, making the binaries to lose too much mass
through the optically thick wind. 
(3)
Binaries beyond the
right boundaries undergo a relatively rapid mass-transfer process due to
the rapid expansion of the MS donors during the HG stage, losing too much mass via the optically thick wind.
(4)
The left boundaries  are determined by
the case that RLOF starts when the MS donor is still in the zero age stage.

In addition, metallicities may have an important effect on the initial contours for producing AIC events.
In Fig.\,2, we set metallicitity $Z=0.02$ (also in Fig.\,4 and Figs\,6$-$7).
If a different metallicity is adopted, the initial contours in Fig.\,2 will be changed.
The initial contours for producing WDs with ${M}_{\rm Ch}$ would be enlarged to have 
larger mass donors and longer orbital periods  with the increase of metallicity
(e.g., Meng et al. 2009; Wang \& Han 2010). 
The influence of metallicity on the initial contours in Fig.\,4 and Figs\,6$-$7 is similar to that of Fig.\,2.
For more studies on the WD+MS channel for producing WDs with ${M}_{\rm Ch}$, see, e.g., 
Li \& van den Heuvel (1997), Hachisu et al. (1999a), Langer et al. (2000), 
Han \& Podsiadlowski (2004, 2006) and Wang et al. (2010), etc.

\subsubsection{Pre-AIC systems}

In the observations, the pre-AIC systems  (i.e., the progenitor systems of AIC events)  with MS donors  
could potentially be identified as supersoft X-ray sources and 
cataclysmic variables (e.g., classical novae, recurrent novae and Ne novae) 
during mass-accretion process.
Supersoft X-ray sources are CO/ONe WD binaries where steady nuclear  burning 
(stable H-/He-shell burning) occurs on the surface of the WDs 
(e.g., van den Heuvel et al. 1992; Rappaport et al. 1994), which are  strong progenitor
candidates of AIC events if the ONe WD can grow in mass close to ${M}_{\rm Ch}$.

Classical novae and recurrent novae are thermonuclear explosions that happen in the surface shell
of accreting CO/ONe WDs (i.e., CO WDs or ONe WDs) in binaries, eventually resulting in  the dynamic 
ejection of the surface shell.  They  usually contain a massive WD with mass-accretion rates below the minimum rate for the
stable shell nuclear burning (see Warner 1995).  
Wu et al. (2017) recently found that nova outbursts can make the WDs growing in mass to ${M}_{\rm Ch}$.
U Sco is a recurrent nova, which contains a 
$1.55\pm0.24\,M_{\odot}$ WD  and a $0.88\pm0.17\,M_{\odot}$ MS donor with an orbit 
period of $\sim0.163$\,d (e.g., Hachisu et al. 2000; Thoroughgood et al. 2001). 
Mason (2011) derived a relatively high [Ne/O] abundance for U Sco during its  outburst in 2010, 
indicating that it may be a nova outburst occurred on the surface of an ONe WD.
However, Mason (2011) neglected the effects of collisions on the formation 
of emission line spectrum (see also Miko\l{}ajewska \& Shara 2017).
Mason (2013) corrected this computational error and 
suggested only a mild overabundance of Ne in the ejecta of U Sco, possibly consistent with a CO WD.

Ne novae are a subclass of classical novae,  
in which an important feature is  the significant enrichment of Ne  detected in the ejecta  
(e.g., Andrea et al. 1994; Gehrz et al. 1998; Denissenkov et al. 2014). 
There are many Ne novae in the observations, which are 
thought to occur on the most massive WDs (e.g., Wanajo et al. 1999; Shore et al. 2003, 2013; Downen et al. 2013). 
It has been estimated that about 1/3 of the observed nova systems contain ONe WDs (see Gil-Pons et al. 2003).
Casanova et al. (2016) recently performed 3D simulations of mixing at the core-envelope interface during nova outbursts. They found
that ONe WDs (as in Ne novae) produce  higher metallicity enhancements in the ejecta than that on the surface of CO WDs 
(i.e., non-Ne novae; see also Glasner et al. 2012). 
The ONe WD usually contains an ONe core surrounded by a  CO-rich mantle, 
where extensive He-shell burning has occurred (e.g., Gil-Pons \& Garc{\'\i}a-Berro 2001; Gil-Pons et al. 2003).

\subsubsection{Post-AIC systems}

After the AIC process, the MS donor may fill its Roche-lobe again, 
and transfer H-rich matter and angular momentum onto the surface of the newborn NS.
The post-AIC systems with MS donors could potentially be identified as
the low-mass X-ray binaries (LMXBs) and the resulting low-mass binary pulsars (LMBPs) in the observations,  
especially the most notably MSPs.
The post-AIC evolution with MS donors is  similar to  that of normal LMXB evolution.
The only difference is that the MS donor has already lost some of its matter during the pre-AIC evolution.
Podsiadlowski et al. (2002) suggested that some of the LMXBs 
consisting a NS and a MS donor with initial orbital periods below the bifurcation period may eventually evolve to 
ultra-compact X-ray binaries (UCXBs) that are a subclass of LMXBs 
with ultra-short orbital periods ($\la$60\,min) and hydrogen-poor donors (see, e.g., 
Nelson 1986; Nelemans \& Jonker 2010; Tauris 2018; Chen et al. 2020).

MSPs are believed to be old radio pulsars with  short spin periods ($<$30\,ms) and 
weak surface magnetic fields ($\sim10^{8}-10^{9}\,{\rm G}$ ),  most of which
are discovered in binaries  (e.g., Lorimer 2008).
In the standard recycling scenario, it has been thought that MSPs can be produced from the evolution of LMXBs, 
in which the NSs originated from core-collapse  SNe  (type Ib/Ic) and 
obtained sufficient mass from their companions, spinning up to millisecond periods finally 
(e.g., Chanmugam \& Brecher 1987; Bhattacharya  \& van den Heuvel 1991).
However,  there is a  discrepancy 
between the rates of Galactic LMXBs and MSPs for the standard recycling scenario (e.g, Kulkarni \& Narayan 1988).
Note that part of the current LMXBs and LMBPs may descend from 
the evolution of intermediate-mass X-ray binaries (IMXBs; e.g., Podsiadlowski et al. 2002; Li 2002, 2015).

As an alternative  way to form MSPs, 
the AIC scenario could possibly help to explain the  rate issue of MSPs in the Galaxy
 (e.g., Bailyn \& Grindlay 1990; Hurley et al. 2010).
Tauris et al. (2013) suggested that ONe WDs with MS donors
could form fully recycled MSPs with wider orbital periods ($\sim10-60$\,d; see also Hurley et al. 2010).
By considering the irradiation-excited wind in accreting ONe WD binaries, 
Ablimit \& Li (2015) found that ONe WDs with MS donors could naturally
reproduce the formation of  LMXBs with the strong-field NSs like  4U 1822$-$37.
They suggested that this channel can explain MSPs with He WD systems for orbital periods in the range of $\sim0.1-30$\,d.
In recent years, with the increasing number of the observed magnetic WDs, 
the magnetic field may play a key role in the evolution of WD binaries (e.g., Ablimit \& Maeda 2019a,b).
Under the assumptions of the magnetic confinement scenario and the evaporative winds, 
Ablimit (2019) recently argued that the AIC scenario
is an alternative way to form  two peculiar kinds of eclipsing MSPs, i.e., redbacks and black widows.
For more studies on the formation of MSPs, see, e.g., Yoon \& Langer (2005), Chen et al.(2013), Jia \& Li (2014), Zhu et al. (2015),
Smedley et al. (2017) and Liu \& Li (2017), etc.

\subsection{The ONe WD+RG channel}

\subsubsection{Evolutionary scenarios}

\begin{figure}
\begin{center}
\epsfig{file=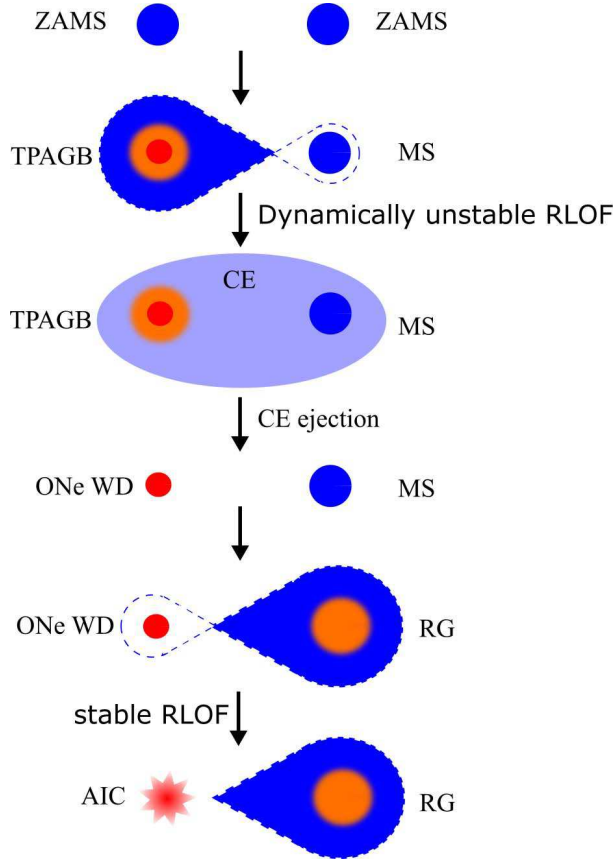,width=8cm} \caption{Evolutionary scenario to ONe WD+RG systems that can form AIC events.}
\end{center}
\end{figure}

In this channel, AIC events originate from the evolution of ONe WD+RG systems.
There is one evolutionary scenario to form ONe WD$+$RG systems and then produce AIC events (see Fig.\,3).
Compared with the ONe WD+MS  channel,
AIC events in the ONe WD$+$RG channel originate from wider primordial binaries  (e.g., Wang 2018a).
The primordial primary fills its Roche-lobe when it evolves to the thermal pulsing AGB (TPAGB) stage. 
In this case, the mass-transfer process is dynamically unstable and a CE may be formed. 
If the CE can be ejected, the primordial primary turns to be an ONe WD. Subsequently, 
the primordial secondary continues to evolve and will become a RG star. 
At this moment, an ONe WD$+$RG system is formed. For this channel, the initial parameters 
of the primordial binaries for producing AIC events are in the range of 
$M_{\rm 1,i}\sim6-8.0\, M_{\odot}$, 
$q<0.3$ 
and $P^{\rm i}\sim1000-6000\,\rm d$.

\subsubsection{Parameter space for AIC events}

\begin{figure}
\begin{center}
\epsfig{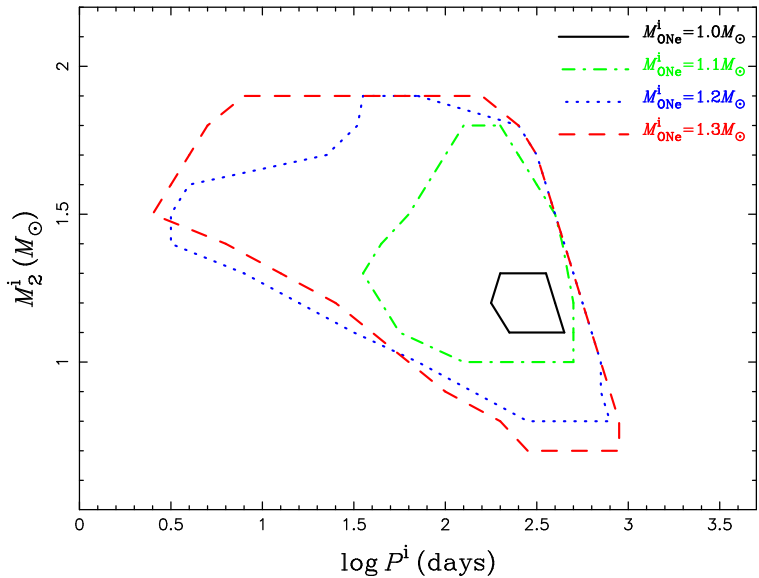} 
\caption{Same as Fig.\,2, but for the ONe WD+RG channel. Source: From Wang (2018a).}
\end{center}
\end{figure}

After the formation of ONe WD+RG systems,  
a CE may be easily formed when the RG star fills its Roche-lobe, resulting in a small parameter space 
for producing WDs with ${M}_{\rm Ch}$  (e.g., Li \& van den Heuvel 1997; Yungelson \& Livio 1998). 
In order to stabilize the mass-transfer process and then to prevent the formation of  the CE, 
Hachisu et al. (1999) argued that the stellar wind from the WDs could strip some mass of the RG stars,  
but this mass-stripping scenario is still under highly debate and has not  been confirmed by the observations (e.g., Ablimit et al. 2014).
By adopting a power-law adiabatic supposition for the mass-transfer prescription
that can be well applicable for the mass-transfer process from a RG donor (e.g., Ge et al. 2010),
Liu et al. (2019) recently enlarged the parameter space of WD+RG systems for producing WDs with ${M}_{\rm Ch}$.
It seems that the adiabatic mass-transfer prescription for RG donor is still highly uncertain (e.g., Woods \& Ivanova 2011),
but the work by Liu et al. (2019)  at least provides an upper limit for the parameter regions that can form WDs with ${M}_{\rm Ch}$. 

Fig.\,4 presents the initial parameter regions of AIC events for the ONe WD+RG channel in 
the $\log P^{\rm i}-M^{\rm i}_2$ plane with different ${M}^{\rm i}_{\rm ONe}$. 
In order to produce AIC events, the ONe WD should have RG companions with initial masses of $\sim0.7-1.9$\,$M_{\odot}$ 
and initial orbital periods of $\sim3-800$\,d. 
The constraints on the boundaries of the initial regions in Fig.\,4 are  similar to that of Fig.\,2.
Note that some previous studies significantly extended the parameter space for forming WDs with ${M}_{\rm Ch}$, 
although their results are strongly dependent on the model assumptions (e.g., L\"{u} et al. 2009; Chen et al. 2011).

\subsubsection{Pre-AIC systems}

The pre-AIC systems with RG donors could potentially be identified as symbiotics in the observations.
Symbiotics play a crucial role in the evolution of semi-detached/detached binaries.
They are CO/ONe WD binaries with long orbital periods, usually consisting 
of a hot WD and a  RG companion (see Truran \& Cameron 1971).
The hot WD could accrete H-rich matter from an evolved RG star through the RLOF process, 
but in most cases through the stellar wind of the RG star. 

Symbiotic novae are a subclass of symbiotics, in which
the mass accretor (CO/ONe WD) experiences a classical nova outburst.
Many symbiotic novae have the WD mass 
close to ${M}_{\rm Ch}$  with RG companions in the observations. 
For example, T CrB  (e.g., Belczy$\acute{\rm n}$ski \& Miko\l{}ajewska 1998), 
RS Oph  (e.g., Brandi et al. 2009; Miko\l{}ajewska \& Shara 2017),  
V745 Sco  (e.g., Drake et al. 2016; Orlando et al. 2017),
and J0757 (e.g., Tang et al. 2012), etc.
Note that it is still not  exclusively identified 
whether the WD in these symbiotic novae is an ONe WD or  a CO WD.

\subsubsection{Post-AIC systems}

Compared with ONe WD+MS systems,
the ONe WD+RG star systems could evolve to form more mildly recycled 
MSPs with He WD companions or even CO WD companions. 
The post-AIC systems with RG donors could potentially be identified as
LMXBs and the resulting young MSPs with long orbital periods ($>$500\,d;  see Tauris et al. 2013).
For the ONe WD+RG star channel, the orbital period at the moment of the AIC and the orbit
for the subsequent post-AIC systems (NS$+$RG systems) are relatively large. 
It has been suggested that such newborn NS binaries with wide separations 
increase the possibility of disruption by stellar encounters in a globular cluster (e.g., Verbunt \& Freire 2014; Belloni et al. 2020),
possibly leading to the formation of  isolated young NSs in globular clusters as suggested by Tauris et al. (2013).

\subsection{The ONe WD+He star channel}

\subsubsection{Evolutionary scenarios}

\begin{figure}
\begin{center}
\epsfig{file=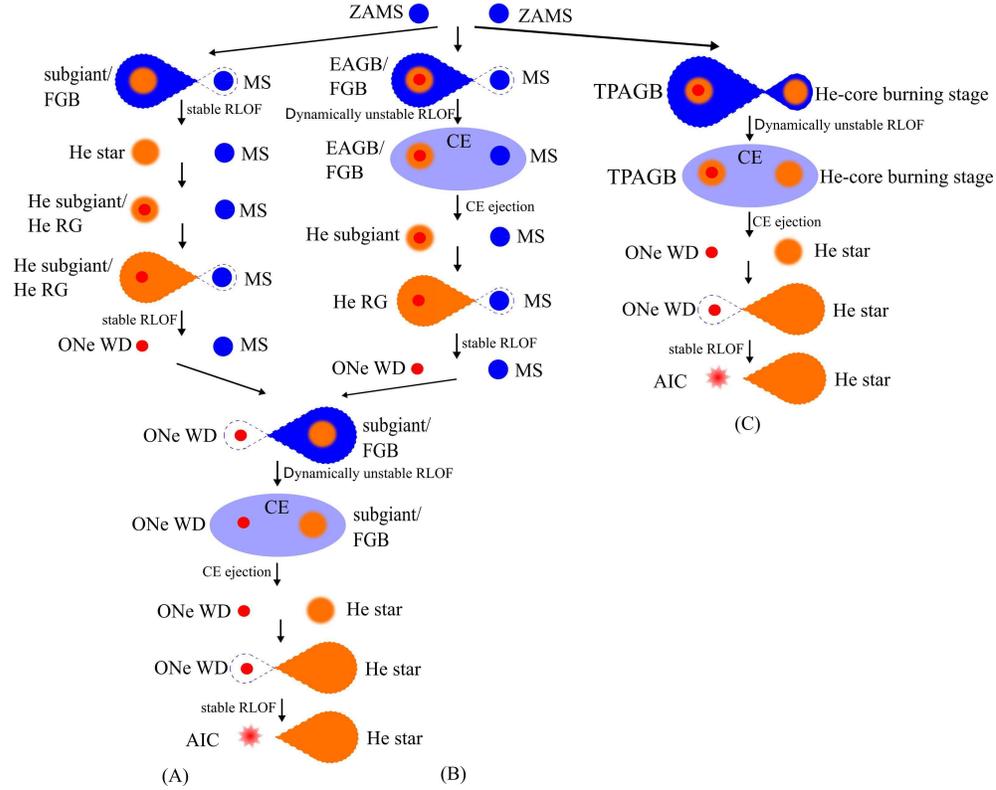,width=13cm} \caption{Evolutionary scenarios to ONe WD+He star systems that can form AIC events.}
\end{center}
\end{figure}

In this channel, AIC events originate from the evolution of ONe WD+He star systems.
According to the evolutionary stage of the primordial primary 
when it first fills its Roche-lobe, there are three evolutionary scenarios to form ONe WD$+$He star systems 
and then produce AIC events (see Fig.\,5). Among the three scenarios, 
AIC events are mainly produced by Scenarios A and B, in which each scenario 
contributes about half of AIC events through the ONe WD+He star channel  (see Wang 2018a). 

\textbf{Scenario A}: 
The primordial primary first fills its Roche-lobe at the HG or FGB branch, 
leading to a stable RLOF process. After the mass-transfer process, the primary turns to be a He star and continues to evolve. 
Subsequently, the He star will exhaust its central He and expand quickly. 
The primary will fill its Roche-lobe again at the He subgiant or He RG stage. 
In this case, the He-rich matter is transferred onto the surface of the secondary MS star in a stable way, 
after which a ONe WD$+$MS system will be formed. The MS secondary continues to evolve and 
will fill its Roche-lobe at the HG or FGB stage. At this stage, the mass-transfer is dynamically unstable and 
a CE may be formed. If the CE can be ejected, an ONe WD$+$He star system is formed.
For this scenario, the initial parameters of the primordial binaries for producing AIC events are in the range of 
$M_{\rm 1,i}\sim6-11\, M_{\odot}$, 
$q>0.2$ 
and $P^{\rm i}<2500\,\rm d$.
More than 40\% of AIC events  from the ONe WD$+$He star channel are produced through this scenario.

\textbf{Scenario B}: 
The primordial primary first fills its Roche-lobe when it evolves to the EAGB or  FGB stage, 
and transfers H-rich matter onto the surface of the secondary in a dynamically unstable way, 
leading to the formation of a CE. After the CE ejection, the primary becomes a He subgiant and continues to evolve. 
The primary will fill its Roche-lobe again at the He RG stage and transfer He-rich matter onto the secondary MS star stably. 
Subsequently, the binary turns to be an ONe WD$+$MS system and 
will evolve to an ONe WD$+$He star system after experiencing the same evolution way as presented in Scenario A.
For this scenario, the initial parameters of the primordial binaries for producing AIC events are in the range of 
$M_{\rm 1,i}\sim6-11\, M_{\odot}$, 
$q>0.4$ 
and $P^{\rm i}\sim300-6000\,\rm d$.
More than half of AIC events from the ONe WD$+$He star channel are produced through this scenario.

\textbf{Scenario C}: 
The primordial primary first fills its Roche-lobe when it evolves to the TPAGB stage and 
the primordial secondary evolves to the He-core burning stage. In this case, the mass-transfer is dynamically unstable 
and a CE with double cores will be formed. If the CE can be ejected, an ONe WD$+$He star system will be produced.
For this scenario, the initial parameters of the primordial binaries for producing AIC events are in the range of 
$M_{\rm 1,i}\sim6-8\, M_{\odot}$, 
$q>0.9$ 
and $P^{\rm i}\sim800-6000\,\rm d$.
Only about 5\% of AIC events from the ONe WD$+$He star channel are produced through this scenario.

\subsubsection{Parameter space for AIC events}

\begin{figure}
\begin{center}
\epsfig{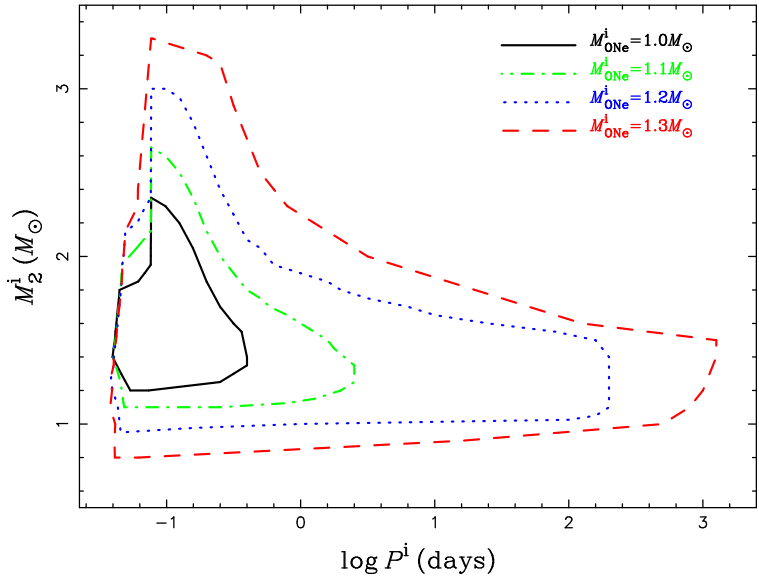} 
\caption{Same as Fig.\,2, but for the ONe WD+He star channel. Source: From Wang (2018a).}
\end{center}
\end{figure}

After the formation of ONe WD+He star systems, the ONe WD could accrete He-rich matter 
from  its He star companion  once it fills its Roche-lobe. 
The accreted He-rich matter is converted into C and O, making the mass growth of the WD.
The ONe WD may form an AIC event when it grows in mass close to ${M}_{\rm Ch}$.
Fig.\,6 represents the initial parameter regions of ONe WD+He star systems 
that can lead to AIC events in the $\log P^{\rm i}-M^{\rm i}_2$ plane with different ${M}^{\rm i}_{\rm ONe}$. 
In order to produce AIC events, the ONe WD should have He star companions with initial masses of $\sim0.8-3.3\,M_{\odot}$ 
and initial orbital periods of $\sim0.04-1000$\,d. 
The constraints on the boundaries of the initial regions in Fig.\,6 are  similar to that of Fig.\,2.

\subsubsection{Pre-AIC systems}

In the observations,  the pre-AIC systems with He star donors could potentially be identified as 
supersoft X-ray sources and  He novae when the ONe WD accretes He-rich matter. 
(1) 
ONe WD+He star systems may appear as supersoft X-ray sources during the stable He-shell burning stage.
The  ONe WD will have a luminosity in the range of  $10^{37}-10^{38}\,{\rm erg\,s^{-1}}$  
when the He-shell burning is stable, consistent with that of the observed supersoft X-ray sources 
($10^{36}-10^{38}\,{\rm erg\,s^{-1}}$;  e.g., Kahabka \& van den Heuvel 1997). 
(2)
V445 Pup is an unique event discovered as  a He nova so far,  in which 
the WD had increased its mass close to  ${M}_{\rm Ch}$  due to the accumulated material  
during nova outbursts (e.g., Ashok \& Banerjee 2003; Kato et al. 2008; Woudt et al. 2009). 
It has been suggested that
V445 Pup contains a CO WD but not an ONe WD due to no enhancement of Ne detected  in the ejecta (see Woudt \& Steeghs 2005). 
Thus, we speculate that V445 Pup may produce an SN Ia but not an AIC event.

\subsubsection{Post-AIC systems}

Compared with the LMBPs, intermediate-mass binary pulsars (IMBPs) include a NS and a massive CO/ONe WD.
The post-AIC systems with He star donors could potentially be identified as IMXBs and 
eventually produce IMBPs with short orbital periods. 
It has been suggested that most of IMBPs may originate from IMXBs that consist of a NS 
and a $2-10\,M_{\odot}$ MS donor, known as the classic IMXB evolutionary scenario 
(e.g.,  Podsiadlowski et al. 2002; Tauris et al. 2012).
However, Tauris et al. (2000)  claimed that  the IMXB evolutionary scenario is hard to form IMBPs with short 
orbital periods ($<$3\,d). Note that the classic IMXB evolutionary scenario may also produce compact IMBPs 
when considering the effect of anomalous magnetic braking of Ap/Bp donors 
(e.g., Justham et al. 2006; Shao \& Li 2012; Liu \& Chen 2014).

Chen \& Liu (2013) proposed an alternative evolutionary way (i.e., the NS+He star channel) 
to form IMBPs with short orbital periods ($<$3\,d), but their work 
can only account for part of the observed IMBPs with short orbital periods (see also Tang et al. 2019). 
Liu et al. (2018a) recently studied the ONe WD+He star channel for the formation of IMBPs. 
They found that this channel may explain most of the observed IMBPs with short orbital periods. 
The studies by Liu et al. (2018a) can well reproduce the observed parameters of 
PSR J1802$-$2124  that is one of the two precisely observed IMBPs.

\subsection{The CO WD+He star channel}

\subsubsection{Parameter space for AIC events}

\begin{figure}
\begin{center}
\epsfig{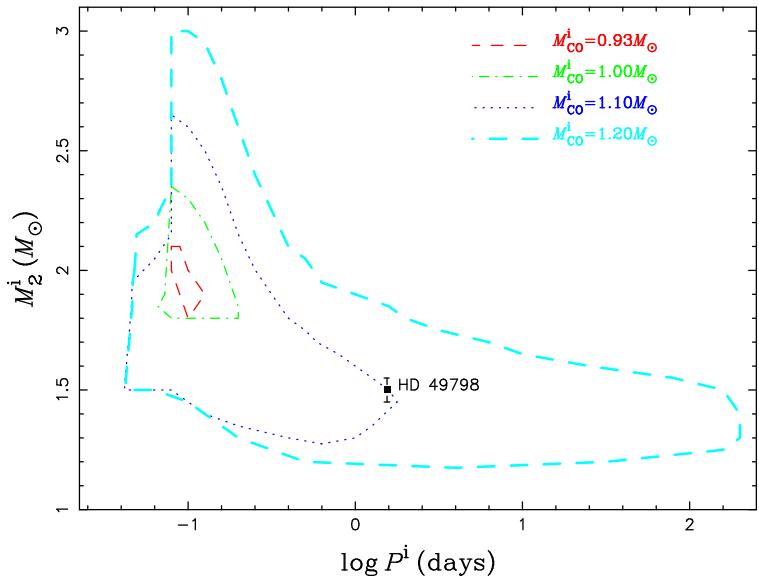} 
\caption{Same as Fig.\,2, but for the CO WD+He star channel.
The square with error bars presents the location of HD 49798 with its massive 
WD companion (see Mereghetti et al. 2009). Source: From Wang (2018a).}
\end{center}
\end{figure}

The CO WD+He star channel has been thought to be one of 
the promising ways to form the observed  SNe Ia in young populations,
in which a CO WD accretes He-rich matter from a He star once it fills its Roche-lobe,
leading to the mass-growth of the WD (e.g., Wang et al. 2009; Ruiter et al. 2009).
Wang et al. (2017) recently found that if the mass-accretion rate is relatively high, 
the He-accreting CO WD will undergo off-center carbon ignition on its surface 
when it grows in mass close to ${M}_{\rm Ch}$  (see also Brooks et al. 2016).
Previous studies usually supposed that the off-center carbon flame will reach 
the center of the CO WD, leading to the formation of an ONe WD 
that will collapse into a NS through the AIC process if the mass-transfer continues 
(e.g., Nomoto \& Iben 1985; Saio \& Nomoto 1985, 1998;  Schwab et al. 2016).  
Note that there are three evolutionary scenarios to form  CO WD+He star systems 
and then to produce  WDs with ${M}_{\rm Ch}$ (for the details see Wang \& Han 2012).

Fig.\,7 presents the initial parameter regions of CO WD+He star systems for producing AIC events
in the $\log P^{\rm i}-M^{\rm i}_2$ plane with different ${M}^{\rm i}_{\rm CO}$.
In this channel, the minimum $M_{\rm CO}^{\rm i}$ for the production of AIC events  is 
$0.93\,M_{\odot}$.
In order to produce AIC events, the CO WD should have He star companions with initial masses of $\sim1.2-3\,M_{\odot}$ 
and initial orbital periods of $\sim0.04-160$\,d. 
The constraints on the upper and left/right-hand boundaries of the initial regions in Fig.\,7 are  similar to that of Fig.\,2.
The regions below the lower boundaries could produce SNe Ia due to 
the center carbon ignition when the WD grows in mass to ${M}_{\rm Ch}$ (see Wang et al. 2017). 
Note that some of the CO WD+He star systems beyond the right-hand boundaries can contribute to 
the formation of massive double CO WDs (e.g., Ruiter et al. 2013; Liu et al. 2018b). 

It is still highly uncertain how the carbon flame propagates when off-center carbon ignition happens
on the surface of CO WDs,
leading to some uncertainties on the final fates of He-accreting CO WDs.
Wu et al. (2020) recently studied the off-center carbon burning of He-accreting 
CO WDs by considering the effect of convective mixing. 
They argued that the temperature of the carbon flame is high enough to burn the carbon into silicon-group elements 
in the outer part of the CO core even though the convective overshooting is considered. They suggested that
the carbon flame will quench somewhere inside the CO WD,  leading to the formation of a COSi WD
that will explode as an SN Ia with Si-rich envelope if the mass accretion continues.

\subsubsection{Pre-AIC systems}

For the CO WD+He star channel, HD 49798 with its massive 
WD companion is a  possible progenitor candidate of AIC events.
HD 49798 is a $1.50\pm0.05\,M_{\odot}$ subdwarf O6 star that has 
a massive compact companion ($1.28\pm0.05\,M_{\odot}$) with an orbital period of
1.548\,d (e.g., Thackeray 1970; Bisscheroux et al. 1997; Mereghetti et al. 2009). 
Krti{\v{c}}ka et al. (2019) recently derived  the mass of  HD 49798 as 
$1.46\pm0.32\,M_{\odot}$ based on  the spectroscopic surface gravity, which
is agreement with the mass obtained from the orbital solution by Mereghetti et al. (2009). 
The binary parameters of HD 49798 with its companion are located in 
the initial contours of CO WD+He star systems for the production of AIC events (see Fig.\,7).
Thus, we speculate that the WD companion of  HD 49798 may collapse into a NS  eventually.
Note that it is still highly uncertain whether  the companion of  HD 49798 is an ONe WD or a CO WD   (e.g., 
Liu et al. 2015; Mereghetti et al. 2016; Popov et al. 2018).

\subsubsection{Post-AIC systems}
CO WD+He star systems could form ONe WD+He star systems  
due to off-center carbon burning happening on the surface of the He-accreting CO WD
if the mass-transfer rate is relatively high (see Wang et al. 2017).
After that, the ONe WD+He star systems continue to evolve and form NS+He star systems if AIC happens, 
in which the newborn NS may be recycled when the He star refills its Roche-lobe again.
Similar to the ONe WD+He star channel, 
the post-AIC systems with He star donors could potentially be identified as IMXBs, 
leading to the formation of  IMBPs finally.

\section{The double-degenerate model}

Close double WDs are the outcomes of low-/intermediate-mass binaries, playing a key role in modern astrophysics. 
They are expected to be potential low-frequency GW emission sources in the Galaxy (see Sect. 4.3)
and contain important information for the constraints on the CE evolution  (e.g., Han 1998; Nelemans et al. 2001a).
In the classic double-degenerate model of AIC events, a WD merges with another WD in a 
compact binary with the total mass heavier than ${M}_{\rm Ch}$, forming an AIC event finally
(e.g., Lyutikov \& Toonen 2017; Ruiter et al. 2019; Liu \& Wang 2020).
The classic double-degenerate model mainly involves the merger of double CO WDs (the double CO WD channel),
the merger of double ONe WDs (the double ONe WD channel), 
and the merger of ONe WD+CO WD systems (the ONe WD+CO WD channel).
In addition, an alternative way could be 
the merging of a WD with a core of an AGB star during the CE evolution, then in a short or a long time leading to AIC
(e.g., Sabach \& Soker 2014; Canals et al. 2018; Soker 2019).

For the mergers of double CO WDs, there are some parameter space to produce AIC events (see Sect. 3.1.2).
For the mergers of double ONe WDs and the ONe WD+CO WD systems, the primary ONe WD 
will collapse into an NS, resulting from  the electron-capture reactions by Mg and Ne  
(e.g., Nomoto \& Iben 1985; Saio \& Nomoto 1985; Wu \& Wang 2018). 
The merging of double WDs may also relate to some high-energy events, 
such as high-energy neutrino emission and short gamma-ray bursts, etc 
(e.g., Xiao et al. 2016; Lyutikov \& Toonen 2017; Rueda et al. 2019). 
Kashyap et al. (2018)  recently argued that the mergers of ONe WD$+$CO WD systems 
might also form the very faint and rapidly fading SNe Ia through a failed detonation.
Liu \& Wang (2020) recently studied the formation of  AIC events through 
the  merging of different kinds of double WDs in a systematic way. 
For more studies on the double-degenerate model of AIC events, 
see, e.g., Lyutikov \& Toonen (2017) and Ruiter et al. (2019).

Aside from the contribution to single NSs via AIC,
the merging of double WDs may also relate to the formation of some other important  peculiar 
objects based on the underlying compositions of the two WDs. For example, 
SNe Ia via the merging of double CO WDs  (e.g., Webbink 1984; Iben \& Tutukov 1984) 
or the merging of CO WD+He-rich WD  (e.g., Dan et al. 2012; Liu et al. 2017; Crocker et al. 2017), 
single hot subdwarfs via the merging of double He WDs (e.g., Han et al. 2003; Zhang \& Jeffery 2012),
R Coronae Borealis (R CrB)  stars and extreme He stars via the merging of CO WD+He WD
(e.g., Webbink 1984; Iben \& Tutukov 1984; Zhang et al. 2014), 
or some Ca-rich transients (i.e., subluminous 2005E-like SNe) via 
the merging of  low-mass CO WD+He WD (e.g., Perets et al. 2010; Meng \& Han 2015), etc.
Note that AM Canum Venaticorum (AM CVn) 
binaries may originate from the evolution of CO WD+He WD systems
(e.g., Warner 1995; Nelemans et al. 2001b) or double He WDs (see Zhang et al. 2018).

\subsection{The double CO WD channel}

\subsubsection{Evolutionary scenarios}

There are three classic binary evolutionary scenarios to form double CO WDs 
that have total mass $\geq$${M}_{\rm Ch}$ and merge in the Hubble time (see Fig.\,11 of Wang 2018b).
For Scenario A, the initial parameters of the primordial binaries for producing AIC events are in the range of
$M_{\rm 1,i}\sim3-9\,M_\odot$, 
$q>0.2$
and $P^{\rm i}<1500\rm d$.
For Scenario B, the initial binary parameters are in the range of
$M_{\rm 1,i}\sim3-7\,M_\odot$, 
$q>0.3$ 
and $P^{\rm i}\sim1300-5000\rm d$.
For Scenario C, the initial binary  parameters  are in the range of
$M_{\rm 1,i}\sim3-7\,M_\odot$, 
$q>0.5$, 
and $P^{\rm i}\sim400-5000\rm d$.
In Scenario C, the double WDs can be also formed after the first CE ejection directly in some cases.
Among the three scenarios, AIC events are mainly produced by Scenarios A and B, in which each scenario 
contributes to about 40\% AIC events  (see Liu \& Wang 2020).
For more discussions on the formation of  double CO WDs, see previous studies 
(e.g.,  Han 1998; Nelemans et al. 2001a; Ruiter et al. 2009, 2013; Toonen et al. 2012; Yungelson \& Kuranov 2017).

The binary evolutionary paths in Fig.\,11 of  Wang (2018b) originate from the CE ejection process before the formation of double CO WDs, 
known as the CE ejection scenarios.
Ruiter et al. (2013)  proposed an important phase to simulate the formation of double CO WDs, in which the primary CO WD is able to
grow in mass by accreting He-rich matter from a He subgaint companion before forming a double CO WD system,  
called the CO WD+He subgiant scenario (see also Liu et al. 2016, 2018b).
Compared with the CE ejection scenario,  the mass-transfer process before the production of double CO WDs
is stable for the CO WD+He subgiant scenario that allows to form massive primary CO WDs and thus
more massive double CO WDs. 
Liu et al. (2018b) recently found that the CO WD+He subgiant scenario has a 
significant contribution to the formation of massive double CO WDs
that may have He-rich atmosphere (double DB/DO WDs in the observations). 
By using SDSS and Gaia data, Genest-Beaulieu \& Bergeron (2019) recently discovered 55 unresolved double DB WDs, which is an 
observational evidence  to support the WD$+$He subgiant scenario for producing massive  double WDs.

\subsubsection{Parameter space for AIC events}
The merging of double CO WDs has been thought to be one of the two major pathways for the formation of SNe Ia
(e.g., Webbink 1984; Iben \& Tutukov 1984).
However, some previous studies suggested that the merging of double CO WDs 
may also lead to the formation of  NSs via the AIC process;
off-center carbon burning may occur on the surface of the massive CO WD 
due to a relatively rapid mass-transfer process during the merging, 
likely converting CO WDs into ONe WDs through an inwardly propagating carbon flame
 (e.g., Nomoto \& Iben 1985; Saio \& Nomoto 1985, 1998;  Timmes et al. 1994).  
 Pakmor et al. (2010) suggested that the AIC may be avoided when the coalescence process is violent, in which
 a prompt detonation is triggered if  the merging continues, resulting in the formation of an SN Ia but not an AIC event
 (known as  the violent merger scenario).
Pakmor et al. (2011) argued that  the mass-ratio of  double CO WDs may have a significant influence on their final  fates. They suggested that
the minimum critical mass-ratio for the violent merger  scenario is $\sim$0.8 (see also Sato et al. 2016). 
This indicates that AIC events are likely to be formed when the mass-ratio of double CO WDs is $<0.8$.

Wu et al. (2019) recently found that 
the outcomes of double CO WDs mainly depend on the merging processes
(e.g., violent merger, fast merger, slow merger and composite merger, etc). 
On the basis of the thick-disc assumption (i.e., slow merger), Wu et al. (2019) found that 
the final evolutionary fates of double CO WDs  are strongly dependent on the mass-accretion rate during the merging, 
which can be divided into four regions: (1) off-centre O/Ne ignitions, then off-centre explosion or Si?Fe cores;
(2) ONe cores, then NSs via AIC; (3) OSi cores, then core-collapse SNe; (4) explosive carbon ignition in the center, then SNe Ia.
They also found that the influence of initial WD mass and cooling time on the final fates of double CO WDs can be ignored.
Compared with the violent merger scenario (e.g., Pakmor et al. 2010, 2011; Sato et al. 2016), the slow merger scenario 
adopted by Wu et al. (2019) allows a low mass-ratio ($<2/3$) for double WDs,  in which the stable mass-transfer can 
occur.\footnote{In the slow merger scenario, the less massive WD will be tidally disrupted and form a pressure-supported disc around the massive WD; 
the mass-accretion rate from the disc onto the massive WD is around the Eddington rate 
and the accreting process can last for millions of years (e.g., Saio \& Jeffery 2000, 2002).}
Note that it is still under highly debate for the outcomes of double CO WDs, 
for more discussions, see,  e.g., Han \& Webbink (1999), Yoon et al. (2007), Chen et al. (2012), Kromer et al. (2013), 
Taubenberger et al. (2013), Moll et al. (2014), Tanikawa et al. (2015),
Fesen et al. (2015) and Bulla et al. (2016), etc.

\subsubsection{Progenitor candidates}

Henize 2-428 and KPD 1930+2752  are two massive WDs that can merge in the Hubble time. 
(1) Henize 2-428 is a bipolar planetary nebula with a double-degenerate core.
Santander-Garc\'ia et al. (2015) suggested that  the nucleus of Henize 2-428 
contains a double CO WD, which has a total mass $\sim$$1.76\,M_\odot$ and mass-ratio $\sim$1 
with an orbital period of $\sim$4.2\,h.  On the basis of the violent merger scenario, the final fate of
Henize 2-428 may be  an SN Ia but not an AIC event.
(2) KPD 1930+2752 is a CO WD+sdB binary,  which has a  total mass  
$\sim1.36-1.48\,M_\odot$ and a $\sim0.45-0.52\,M_\odot$  sdB star 
with  an   orbital period of $\sim$2.28\,h (e.g., Maxted et al. 2000; Geier et al. 2007).
Liu et al. (2018b) recently suggested that 
KPD 1930+2752 will form a double WD in $\sim$200\,Myr due to the evolution of the sdB star, 
after which it will merge in $\sim$4\,Myr. The final fate of
KPD 1930+2752 may be an AIC event  as its mass-ratio is far below 0.8.

Meanwhile, there are  some other massive
double WDs in the observations that may have the total mass close to ${M}_{\rm Ch}$, e.g.,  
GD687  (e.g., Geier et al. 2010), 
SBS 1150+599A  (see Tovmassian et al. 2010),
V458 Vulpeculae  (see Rodr\'{\i}guez-Gil et al. 2010),
WD 2020-425  (see Napiwotzki et al. 2007),
 and NLTT 12758 (Kawka et al. 2017), etc.  
Hollands et al. (2020) recently reported the identification of a  1.14\,$M_\odot$ WD, i.e., WD J0551+4135. 
They argued that the ultra-massive WD with an unique hydrogen/carbon mixed atmosphere 
is consistent with the merger remnant of double WDs in a tight binary.
In addition, Borges et al. (2020) argued that the anomalous X-ray pulsar 4U 0142+61 
may harbor a fast-rotation magnetic young  WD with mass near ${M}_{\rm Ch}$,
which can be the recent outcome of the merging from two less massive WDs.
It has been proposed that the continuous GW emission may 
be one of the probes to detect the ultra-massive WDs directly by the future various space-based detectors (see Kalita et al. 2020). 

Some systematic surveys have been proposed to search for double WDs,
e.g.,  the SWARMS survey  (e.g., Badenes et al. 2009) and the ESO SN~Ia Progenitor Survey
(SPY; e.g., Koester et al. 2001; Napiwotzki et al. 2004; Nelemans et al. 2005; Geier et al. 2007).
Carrasco et al. (2014)  estimated that 
a large number of double WDs may be discovered by Gaia
 (see also Toonen et al. 2017). 
Tian et al. (2020a)  recently provide hundreds of double WDs and more than ten thousand WD+MS binaries selected from Gaia DR2. 
In addition, the GPS1 (see Tian et al. 2017) and its extended  proper motion catalogues (GPS1+; see Tian et al. 2020b)
could provide a more substantial population of WD binaries due to its accurate photometric and astrometric 
data from Gaia DR1 (DR2), PS1, SDSS and 2MASS.

\subsection{The double ONe WD channel}

\begin{figure}
\begin{center}
\epsfig{file=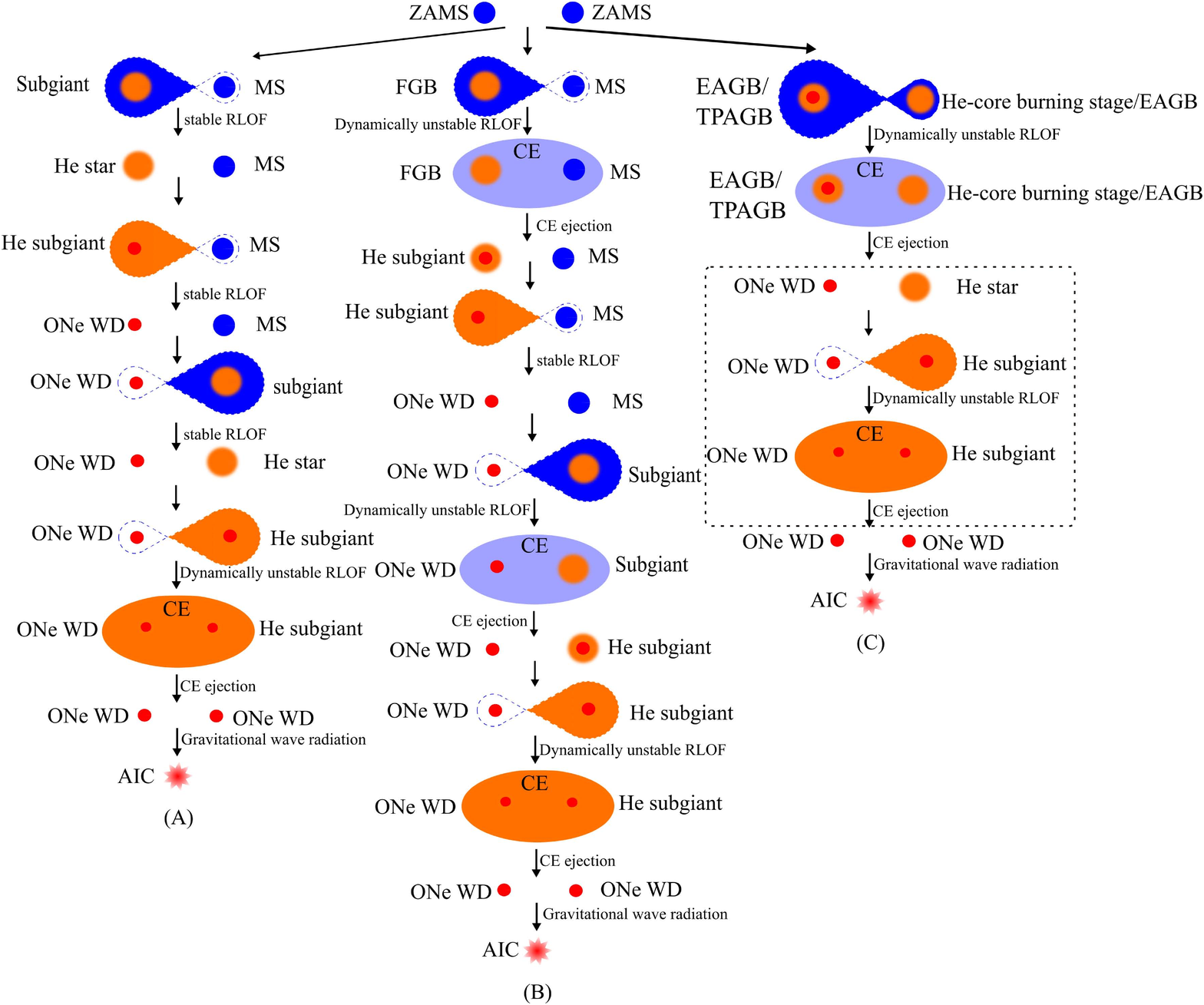,width=14cm} \caption{Evolutionary scenarios to double ONe WDs that can form AIC events.
In Scenario C, some cases will not experience the second CE ejection (dashed box).}
\end{center}
\end{figure}

According to the evolutionary stage of the primordial primary when it first fills its Roche-lobe, 
there are three evolutionary scenarios to form double ONe WDs that can produce AIC events (see Fig.\,8),
in which Scenario A is the main way to form AIC events  (see Liu \& Wang 2020).

\textbf{Scenario A}: 
The primordial primary first fills its Roche-lobe at the HG stage and transfers H-rich matter stably onto the secondary. 
After the mass-transfer process, the primary turns to be a He star and continues to evolve. 
The He star will fill its Roche-lobe again when its central He is exhausted and evolve to be a He subgiant. 
In this case, the He subgiant transfers He-rich matter onto the MS secondary stably, 
after which the primary turns to be an ONe WD. Subsequently, the primordial secondary 
will fill its Roche-lobe at the HG stage and start a stable RLOF process. 
After that, the secondary evolves to a He star and continues to evolve. 
The secondary would fill its Roche-lobe again when its central He is exhausted and evolve to a He subgiant. 
In this case, the mass-transfer is dynamically unstable and a CE may be formed. 
After the CE ejection, the secondary become another ONe WD, i.e., a double ONe WD system is formed.
For this scenario, the initial parameters of the primordial binaries for producing AIC events are in the range of 
$M_{\rm 1,i}\sim8-11\, M_{\odot}$, 
$q\sim0.2-0.8$ 
and $P^{\rm i}<40\,\rm d$.
About 78\% of AIC events from the double ONe WD merger channel are produced through this scenario.

\textbf{Scenario B}: 
The primordial primary first fills its Roche-lobe at the FGB phase, 
and transfers H-rich matter onto the surface of the secondary. 
In this case, the mass-transfer process is dynamically unstable and a CE would be formed. 
If the CE can be ejected, the primary turns to be a He subgiant. 
The He subgiant will fill its Roche-lobe again quickly and transfer He-rich matter onto the MS secondary stably. 
After the mass-transfer process, the binary becomes an ONe WD$+$MS system. 
Subsequently, the primordial secondary will evolve to the HG stage and fill its Roche-lobe. 
In this case, the mass-transfer is dynamically unstable and a CE may be formed. 
After the CE ejection, the secondary becomes a He subgiant and will fill its Roche-lobe again. 
At this stage, the mass-transfer is also dynamically unstable and a CE will be formed. 
If the CE can be ejected, a double ONe WD system will be produced.
For this scenario, the initial parameters of the primordial binaries for producing AIC events are in the range of 
$M_{\rm 1,i}\sim8-11\, M_{\odot}$, 
$q>0.7$ 
and $P^{\rm i}\sim500-1000\,\rm d$.
About 7\% of AIC events from the double ONe WD merger channel are produced through this scenario.

\textbf{Scenario C}: 
The primordial primary fills its Roche-lobe when it evolves to the EAGB or TPAGB phase 
and the primordial secondary evolves to the He-core burning stage or the EAGB stage. 
In this case, a CE may be formed due to the dynamically unstable mass-transfer. 
After the CE ejection, an ONe WD$+$He star system will be produced. 
The He star continues to evolve and will evolve to a He subgiant when its central He is exhausted. 
At this stage, the He subgiant will expand quickly and fill its Roche-lobe. 
The second CE may be formed because of dynamically unstable RLOF. 
If the CE can be ejected, a double ONe WD system will be produced.
Note that the second CE evolution process shown in 
the dashed box in Fig.\,8 may be not indispensable in some cases.
For this scenario, the initial parameters of the primordial binaries for 
producing AIC events are in the range of 
$M_{\rm 1,i}\sim6-9\, M_{\odot}$, 
$q>0.8$ 
and $P^{\rm i}\sim400-6000\,\rm d$.
About 15\% of AIC events from the double ONe WD channel are produced through this scenario.

\subsection{The ONe WD+CO WD channel}

\begin{figure}
\begin{center}
\epsfig{file=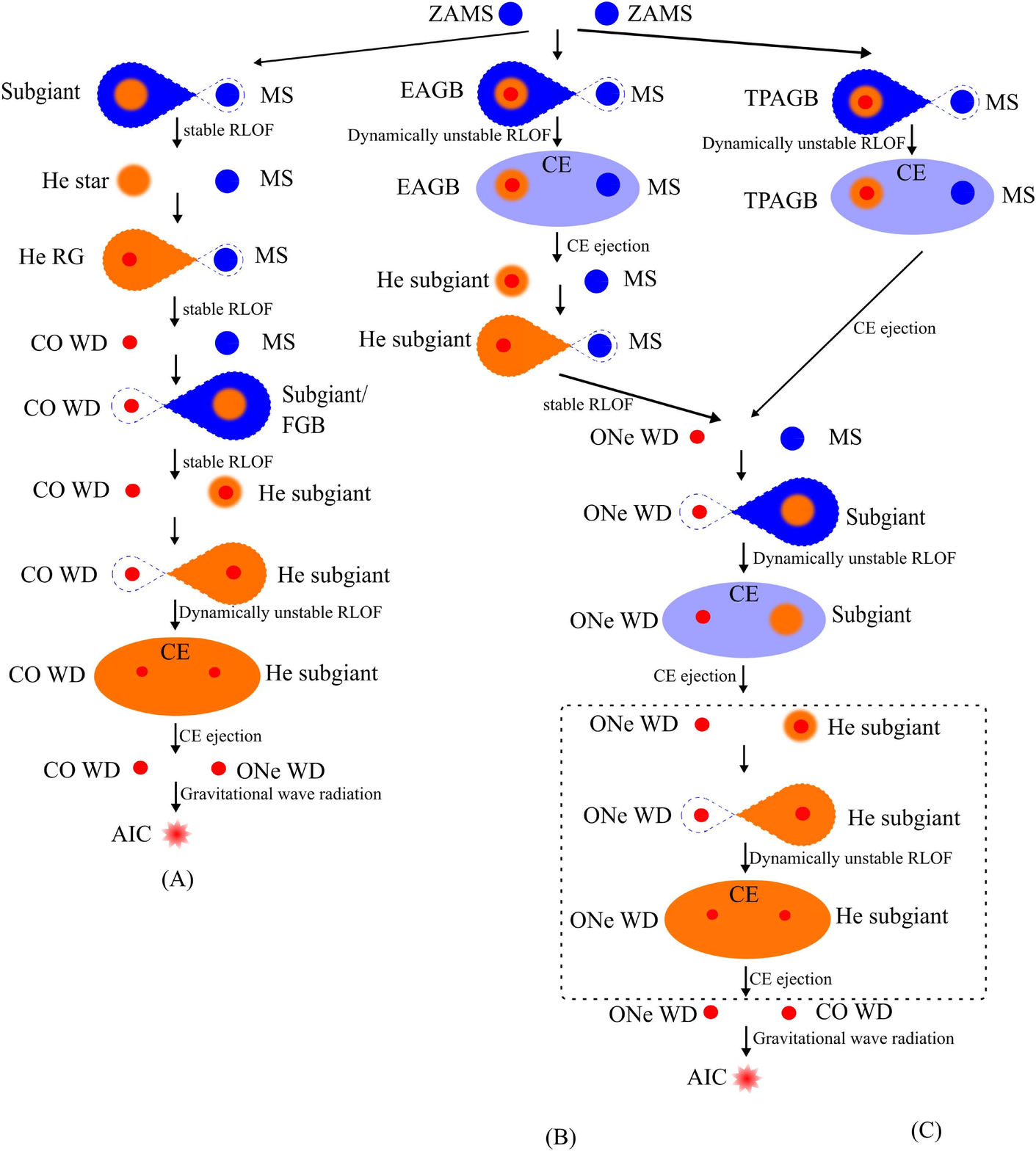,width=13cm} \caption{Evolutionary scenarios to ONe WD+CO WD systems that can form AIC events.
In Scenarios B and C, some cases will not experience the third CE ejection (dashed box).}
\end{center}
\end{figure}

According to the evolutionary stage of the primordial primary when it first fills its Roche-lobe, 
there are three evolutionary scenarios to form ONe WD$+$CO WD systems and then merge to produce AIC events  (see Fig.\,9), 
in which Scenario A is the main way to form AIC events  (see Liu \& Wang 2020).
In Scenario A, the primordial primary evolves to a CO WD first and then the primordial secondary forms an ONe WD,
whereas it will be the other way around for Scenarios B and C.

\textbf{Scenario A}: 
The primordial primary first fills its Roche-lobe at the HG stage and 
transfers H-rich matter onto the surface of the primordial secondary stably, 
leading to a mass reverse of the binary, i.e., an algol binary is formed. 
After the mass-transfer process, the primordial primary turns to be a He star and continues to evolve. 
The He star will exhaust its central He and evolve to a He RG. 
At this stage, the He star will expand quickly and fill its Roche-lobe again. 
The primordial primary transfers He-rich matter onto the secondary stably. 
After the stable RLOF, the binary turns to be a CO WD$+$MS system. 
Subsequently, the primordial secondary fills its Roche-lobe at the HG or FGB stage and experiences a stable RLOF process. 
After that, the primordial secondary becomes a He subgiant and will fill its Roche-lobe again. 
In this case, a CE may be formed due to the dynamically unstable RLOF. 
If the CE can be ejected, an ONe WD$+$CO WD system will be produced.
For this scenario, the initial parameters of the primordial binaries for producing AIC events are in the range of 
$M_{\rm 1,i}\sim4-11\, M_{\odot}$, 
$q>0.2$ 
and $P^{\rm i}<1000\,\rm d$.
About half of AIC events from the ONe WD$+$CO WD merger channel are produced through this scenario.

\textbf{Scenario B}: The primordial primary fills its Roche-lobe when it evolves to the EAGB phase, 
leading to the formation of a CE due to the dynamically unstable RLOF. 
If the CE can be ejected, the primordial primary becomes a He subgiant and continues to evolve. 
The He subgiant will fill its Roche-lobe quickly and transfer He-rich matter onto the MS secondary stably. 
After the mass-transfer process, the binary turns to be an ONe WD$+$MS system. 
Subsequently, the secondary will fill its Roche-lobe at the HG phase, 
and transfer H-rich matter onto the surface of the ONe WD. 
In this case, the mass-transfer is dynamically unstable and a CE may be formed. 
After the CE ejection, an ONe WD$+$He subgiant system will be produced. 
The He subgaint will quickly fill its Roche-lobe again, and transfer He-rich matter onto the ONe WD. 
At this stage, the second CE may be formed due to the dynamically unstable RLOF. 
If the CE can be ejected, an ONe WD$+$CO WD system will be formed finally.
Note that the third CE evolution process shown in the dashed box in Fig.\,9 may be not indispensable in some cases.
For this scenario, the initial parameters of the primordial binaries for producing AIC events are in the range of 
$M_{\rm 1,i}\sim2-11\, M_{\odot}$, 
$q>0.2$ 
and $P^{\rm i}\sim400-6000\,\rm d$.
About 30\% of AIC events from the ONe WD$+$CO WD merger channel are produced through this scenario.

\textbf{Scenario C}: 
The primordial primary first fills its Roche-lobe when it evolves to the TPAGB phase. 
In this case, the mass-transfer is dynamically unstable and a CE may be formed. 
After the CE ejection, an ONe WD$+$MS system is produced. 
Subsequently, the binary will evolve to an ONe WD$+$CO WD system 
after experiencing similar evolution as presented in Scenario B.
For this scenario, the initial parameters of the primordial binaries 
for producing AIC events are in the range of 
$M_{\rm 1,i}\sim5-9\, M_{\odot}$, 
$q>0.3$ 
and $P^{\rm i}\sim200-1600\,\rm d$.
About 20\% of AIC events from the ONe WD$+$CO WD merger 
channel are produced through this scenario.

\section{Binary population synthesis}

Binary population synthesis is a useful approach to simulate a large population of stars or binaries, 
especially for the formation of peculiar stars,
and then to compare the theoretical results with those of observations 
(e.g., Han et al. 1995; Yungelson \& Livio 1998; Nelemans et al. 2001a,b; Hurley et al. 2010).
Table\,1 presents the estimated Galactic rates, NS numbers and delay times  of AIC events 
from various channels with different CE ejection parameters.
The estimated Galactic rates of AIC events are strongly dependent on the value of the CE ejection 
parameter $\alpha_{\rm CE}\lambda$ that is still highly uncertain (e.g., Ivanova et al. 2013).

\begin{table}
 \begin{center}
 \caption{The estimated Galactic rates, NS numbers and delay times 
 of AIC events for various channels with different values of CE ejection parameters,  in which
 we adopt metallicity $Z=0.02$ and a constant
star-formation rate of $5\,{M}_{\odot}\rm yr^{-1}$ in our Galaxy. 
 The results for different single-degenerate models are from Wang (2018a), 
 whereas the results for different double-degenerate models are from Liu \& Wang (2020).
 Notes: 
 $\alpha_{\rm CE}\lambda$ = CE ejection parameter; 
$\nu_{\rm AIC}$ = Galactic rates of AIC events;
$\rm Number$ = Expected  number of  NS systems from the single-degenerate model 
or  single isolated NSs from the double-degenerate model  in the Galaxy.}
   \begin{tabular}{cccccccccccccc}
\hline \hline
$\rm Channels $ & $\alpha_{\rm CE}\lambda$  & $\nu_{\rm AIC}$  & $\rm Number$ & ${\rm Delay\,Times}$\\
& &($10^{-3}$\,yr$^{-1}$) & $(10^{\rm 7})$ &(Myr)\\
\hline
$\rm ONe\,WD$+$\rm MS$          & $0.5$    & $0.138$      & $0.165$   & $110-1400$ \\
$\rm ONe\,WD$+$\rm MS$          & $1.5$    & $0.079$      & $0.095$   & $70-1400$ \\
\hline
$\rm ONe\,WD$+$\rm RG$          & $0.5$    & $0.012$      & $0.014$  & $1400-6300$ \\
$\rm ONe\,WD$+$\rm RG$          & $1.5$    & $0.033$      & $0.040$  & $1400-6300$\\
\hline
$\rm ONe\,WD$+$\rm He\,star$   & $0.5$    & $0.105$      & $0.126$  & $40-140$ \\
$\rm ONe\,WD$+$\rm He\,star$   & $1.5$    & $0.676$      & $0.811$  & $30-180$ \\
\hline
$\rm CO\,WD$+$\rm He\,star$     & $0.5$    & $0.083$      & $0.100$ & $50-110$ \\
$\rm CO\,WD$+$\rm He\,star$     & $1.5$    & $0.129$      & $0.155$ & $50-110$ \\
\hline
$\rm Double\,CO\,WDs$              & $0.5$      & $1.129$         & $1.356$ & $>90$\\
$\rm Double\,CO\,WDs$              & $1.5$      & $5.160$         & $6.194$  & $>110$\\
\hline
$\rm Double\,ONe\,WDs$            & $0.5$      & $0.051$         & $0.061$ & $>55$\\
$\rm Double\,ONe\,WDs$            & $1.5$      & $0.285$         & $0.342$ & $>55$\\
\hline
$\rm ONe\,WD$+$\rm CO\,WD$ & $0.5$    & $0.268$      & $0.321$ & $>55$\\
$\rm ONe\,WD$+$\rm CO\,WD$ & $1.5$    & $3.486$      & $4.184$ & $>55$\\
\hline
\end{tabular}
\end{center}
\end{table}

\subsection{Predicted rates of AIC events}

\begin{figure}
\begin{center}
\epsfig{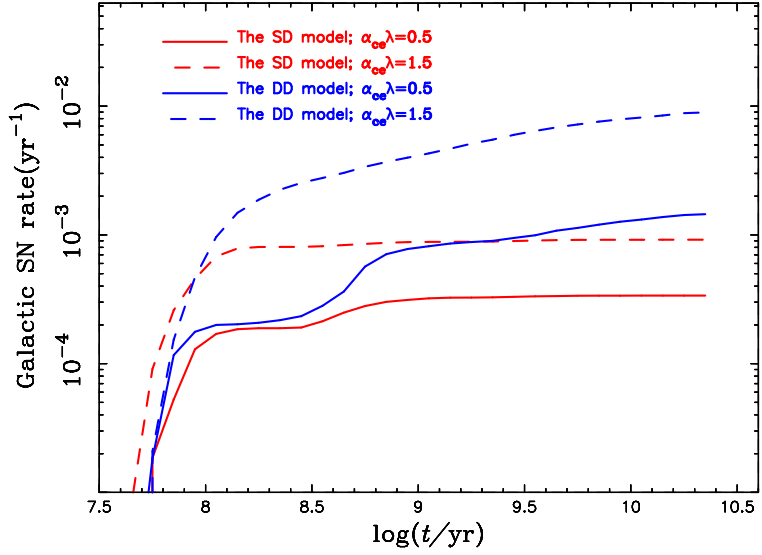}
\caption{Evolution of the predicted AIC rates as a function of time in the Galaxy 
for the single-degenerate (SD) and double-degenerate (DD) models 
with  different $\alpha_{\rm CE}\lambda$ by adopting metallicity $Z=0.02$ and a constant
star-formation rate of $5\,{M}_{\odot}\rm yr^{-1}$. 
The results for the SD model are from Wang (2018a), 
whereas the results for the DD model are from Liu \& Wang (2020).}
\end{center}
\end{figure}

Fig.\,10 shows the evolution of AIC rates changing with time in the Galaxy. 
The estimated rates of AIC events in the Galaxy are in the range of
$\sim0.3-0.9\times10^{-3}\,\rm yr^{-1}$ for the single-degenerate model and 
$\sim1.4-8.9\times10^{\rm -3}\,\rm yr^{\rm -1}$ for the double-degenerate model, in which
the double-degenerate model plays a dominant way.
Fryer et al. (1999) estimated that the rates of  AIC events ranges from $10^{-7}-10^{-5}\,\rm yr^{-1}$ 
by modelling the  r-process nucleosynthetic yields  of neutron-rich ejecta through AIC process.
Although it is still under debate whether the AIC process produces the r-process elements or not 
 (e.g., Fryer et al. 1999; Qian \& Wasserburg 2007),
the theoretical AIC rates in this review at least provide an upper limit for the AIC events.

The estimated Galactic number of NS systems originating from the single-degenerate 
model may be in the range of $\sim0.4-1.1\times10^{\rm 7}$, 
and the single NS number  from the double-degenerate model ranges from $\sim0.2-1.1\times10^{\rm 8}$.
In the single-degenerate model, the ONe WD+He star channel is the main way to produce AIC events, 
and we cannot ignore the contribution of the CO WD+He star channel when studying AIC events.
In the double-degenerate model,  the double\,CO\,WD channel plays a main way for the formation of AIC events though there exist
some uncertainties for this channel.  If we did not consider the contribution of double CO WD mergers,
the  Galactic rates of AIC events from the double-degenerate model  will decrease to 
$\sim0.6-4.7\times10^{\rm -3}\,\rm yr^{\rm -1}$, and the  corresponding single NS number decreases to 
$\sim0.4-4.5\times10^{\rm 7}$.

Lyutikov \& Toonen (2017) recently investigated the formation of AIC events 
from the merging of  ONe WD$+$CO WD systems in a systematic way. 
The predicted rates of AIC events in Table 1 for the ONe WD$+$CO WD 
channel is compatible to that of Lyutikov \& Toonen (2017) if a small value of $\alpha_{\rm CE}\lambda$ is adopted.
Using a  binary population synthesis method,
Ruiter et al. (2019) recently systematically studied  the single-/double-degenerate  models of AIC events. 
The predicted Galactic rates of AIC events in this review are somewhat higher than those of 
Ruiter et al. (2019). The main reason is that Ruiter et al. (2019) did not consider the contribution of 
the CO WD+He star channel in the single-degenerate model and 
the double\,CO\,WD channel in the double-degenerate model.

\begin{figure}
\begin{center}
\epsfig{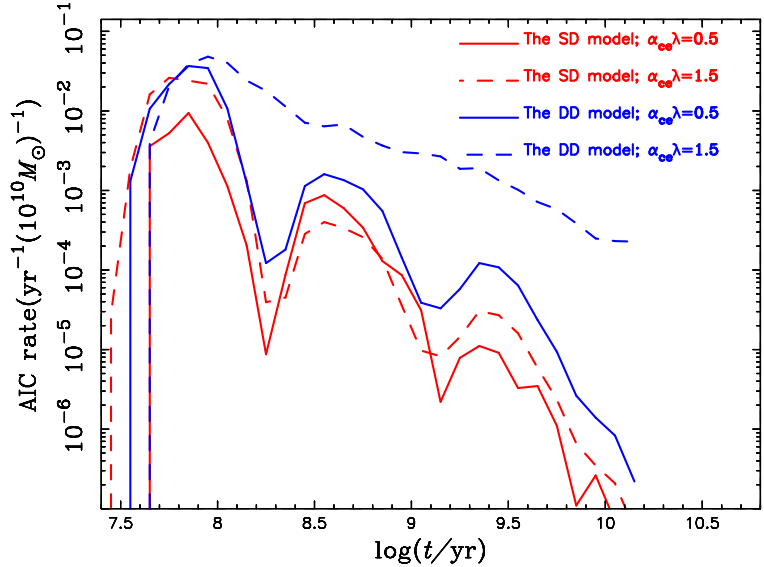}
\caption{Delay time distributions  of AIC events based on a single starburst with a total mass of $10^{10}\,M_{\odot}$. 
The results for the SD model are from Wang (2018a), 
 whereas the results for the DD model are from Liu \& Wang (2020).}
\end{center}
\end{figure}

The delay times  of AIC events represent the time interval from the star formation to the occurrence of AIC. 
Fig.\,11 represents the delay time distributions of AIC events based on a single starburst with a total
mass of $10^{10}\,M_{\odot}$ in stars. In the single-degenerate model, 
the ONe/CO WD+He star channels mainly contribute to AIC events with short delay times after the starburst,
the ONe WD+MS channel for the intermediate delay times, and
the ONe WD+RG channel for the long delay times. 
AIC events with the shortest delay times have a He star donor at the moment of 
NS formation, and the corresponding post-AIC systems with He star donors may eventually 
produce IMBPs with short orbital periods (see Sects 2.3.4 and 2.4.3).
AIC events with the longest delay times have a RG donor at the moment of NS formation, 
and most of  the post-AIC systems with RG donors will form NS+He WD systems with long orbital periods  finally, 
which may be identified as young MSPs in globular clusters (see Sect. 2.2.4).
The double-degenerate model will produce AIC events from 55\,Myr to the Hubble time, which
is determined by the merging timescale of double WDs resulted from the GW emission. 
The minimum delay time in Fig.\,11 is mainly determined by the MS lifetime of  
the maximum  mass  of a star  forming  an ONe WD, depending on the change of metallicities (e.g., Doherty et al. 2017).

\subsection{Mass distribution of NSs }

\begin{figure*}
\begin{center}
\epsfig{file=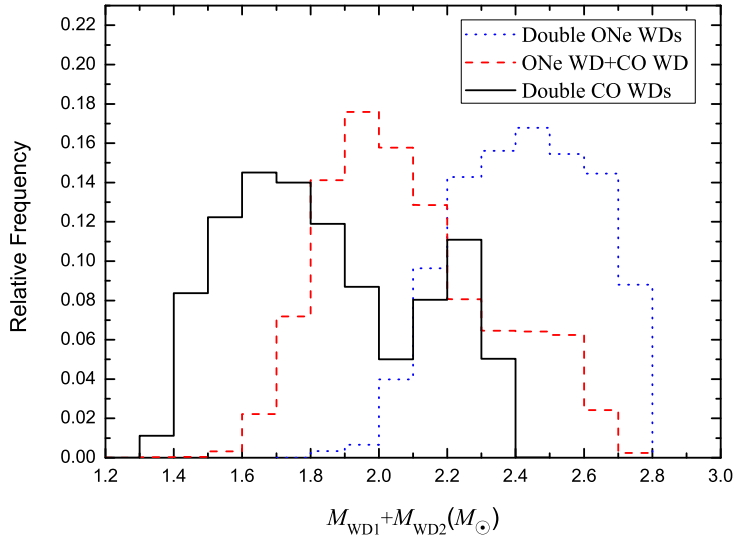,angle=0,width=12.cm}
\end{center}
\caption{The total mass distributions of different kinds of double WD mergers for producing AIC events
with $\alpha_{\rm CE}\lambda=1.0$. 
The data points forming these distributions  are from Liu \& Wang (2020).} 
\end{figure*}

The post-AIC events from the single-degenerate model could
potentially be identified as low/intermediate-mass X-ray binary pulsars, and 
then the resulting low/intermediate-mass binary pulsars that may have 
NSs with masses in the range of $\sim1.25-1.7\,M_{\odot}$ (e.g., Tauris et al. 2013; Liu et al. 2018).
However, the NSs resulted from the double-degenerate model are quite different from that of the single-degenerate model. 
The merging of double WDs will form single isolated NSs with a large mass range after the AIC process.
Comparing with the single NSs originating from the classic core-collapse SNe,
the single NSs from the mergers of double WDs  may correspond to a specific kind of NSs, 
showing some different  properties, such as circumstellar environments and magnetic fields, etc. 
Beznogov et al. (2020)  recently suggested that the NSs originated from AIC
may be observable during the neo-NS (i.e., hot NS) stage, in which the NS has just become transparent to neutrinos.
For more studies on the isolated NSs, see, e.g., Tauris \& van den Heuvel (2006), Popov (2011),  
Liu \& Li (2019), Soker (2020) and Jiang et al. (2020), etc.

Fig.\,12 presents the total mass distribution of different double WD mergers for producing AIC events.
The total masses of double WD mergers are mainly in the range of $\sim1.4-2.8\,M_{\odot}$, 
which at least provides an upper limit for the  maximum mass of the formed single isolated 
NSs.  Currently, the most reliable constraints on the maximum mass of NSs are from
the mass determinations of the massive pulsar binaries in the observations, 
e.g.,  the PSR~J0740+6620 with $M_{\rm NS}=2.14_{-0.09}^{+0.10}\,M_{\odot}$ (see Cromartie et al. 2020),
the PSR~J0348+0432 with $M_{\rm NS} = 2.01\pm0.04\,M_{\odot}$ (see Antoniadis et al. 2013),
and the PSR~J1614-2230 with $M_{\rm NS} = 1.97\pm0.04\,M_{\odot}$ (Demorest et al. 2010). 
These observational values provide a lower limit on the maximum mass of NSs.

For the double-degenerate model of AIC events,
the total merging masses are mainly in the range of 
$1.4-2.4\,M_{\odot}$ for the mergers of double CO WDs, 
$2.0-2.8\,M_{\odot}$ for the mergers of double ONe WDs,
and $1.6-2.7\,M_{\odot}$ for the mergers of ONe WD$+$CO WD systems.
The total  mass distributions in Fig.\,12 have two peaks for the double CO WD mergers, in which the 
left peak mainly originates from the classic CE ejection scenarios (see Fig. 11 of Wang 2018b),
and the right peak mainly originates from the CO WD$+$He subgiant scenario that can make the primary CO WDs  
growing in mass up to $0.48\,M_{\odot}$ after its formation (see Liu et al. 2018b).

\subsection{Gravitational wave signals}

\begin{figure}
\begin{center}
\begin{tabular}{@{}c@{}}
\centerline{\epsfig{file=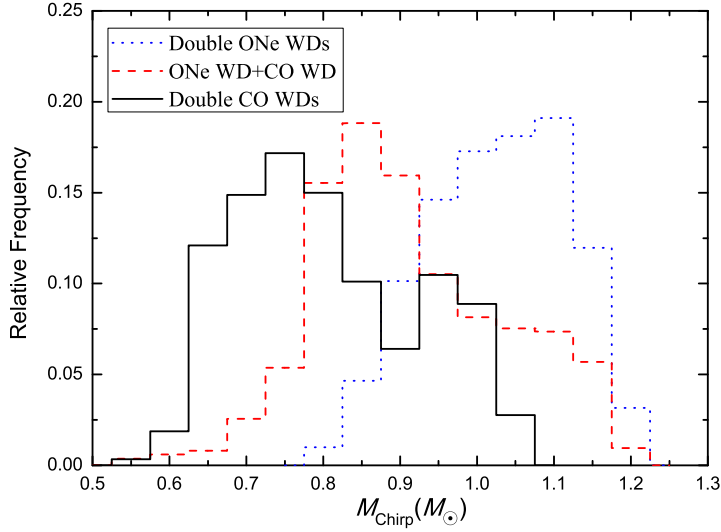,angle=0,width=12.cm}}
\end{tabular}
 \caption{Chirp mass distributions of different kinds of double WDs for producing 
 AIC events with $\alpha_{\rm CE}\lambda=1.0$. 
The data points forming these distributions are from Liu \& Wang (2020). }
\end{center}
\end{figure}

After the detection of the first double black hole merger by the ground-based aLIGO/Virgo,
many double black hole mergers and 1 double NS merger have been confirmed in the past few years, 
starting a new era of GW  astronomy (e.g., Abbott et al. 2016, 2017a,b).
Close double WDs with short orbital periods are expected to dominate the Galactic GW 
background in the range of $10^{-4}-10^{-1}$\,Hz  
(e.g., Evans et al. 1987; Nelemans et al. 2001a).
They would be observable 
by the future space-based GW detectors, such as the Laser Interferometer Space 
Antenna (LISA; e.g., Nelemans et al. 2004; Ruiter et al. 2010; Marsh 2011),
TianQin  (e.g., Luo et al. 2016; Wang et al. 2019; Bao et al. 2019;
Shi et al. 2019; Feng et al. 2019) 
and Taiji (e.g., Ruan et al. 2019, 2020a,b; Luo et al. 2020), etc. 
Nelemans (2013)  estimated that several thousand double WDs should be individually detected by LISA.
Kremer et al. (2017) recently argued that about 2700 double WDs 
will be detected by  the space-based GW detectors like LISA.
Fore more studies about the GW signals from double WDs, see, e.g, 
Liu (2009), Yu \& Jeffery (2010, 2015), 
Liu et al. (2012), Liu \& Zhang (2014) and Zou et al. (2020), etc.

The chirp mass is a mass function of double WDs that can be measured by the GW detectors directly. 
Fig.\,13 shows the chirp mass distributions of different kinds of  double WDs for producing AIC events at the moment of their birth,
which mainly ranges from 0.55 to $1.25\,M_{\odot}$. 
For the case of double CO WDs, their chirp masses are in the range of $0.55-1.05\,M_{\odot}$ 
and there are two peaks at $0.75\, M_{\odot}$ and $0.95\, M_{\odot}$, in which the origin of the double peaks are similar to that of Fig.\,12.
The chirp masses of ONe WD$+$CO WD systems are in the range of $0.55-1.2\, M_{\odot}$ and have a peak at $0.85\, M_{\odot}$, 
whereas the chirp masses of double ONe WDs range from 0.8 to $1.25\,M_{\odot}$ and have a peak at $1.1\,M_{\odot}$.

\begin{figure}
\begin{center}
\begin{tabular}{@{}c@{}}
\centerline{\epsfig{file=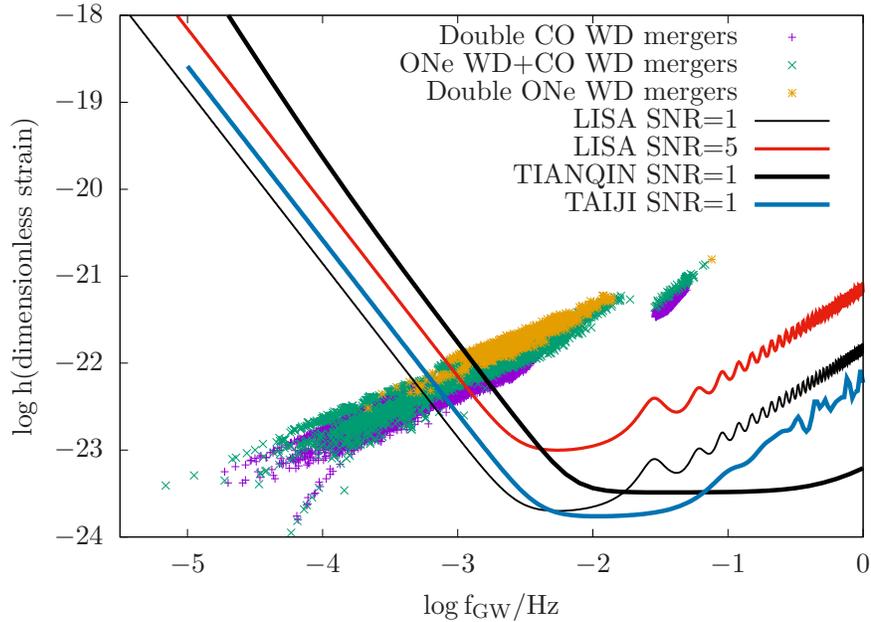,angle=0,width=14.cm}}
\end{tabular}
 \caption{Dimensionless GW strain amplitude of different kinds of double WDs for 
producing AIC events with $\alpha_{\rm CE}\lambda=1.0$, in which 
we simply set the distance from double WDs to the detectors as 
8.5\,kpc. The sensitivity curves for the future space-based 
GW observatory LISA is based on Larson, Hiscock \& Hellings: http://www.srl.caltech.edu/~shane/sensitivity. 
The sensitivity curve for TianQin is from Wang et al. (2019).  
The sensitivity curve for Taiji is from Ruan et al. (2020b).
The data points forming these distributions are from Liu \& Wang (2020). }
\end{center}
\end{figure}

The GW strain amplitude ($h$) is defined as the fractional change in separation once a GW passes through the detectors.
Fig.\,14 presents the dimensionless $h$ distributions of different kinds of double WDs for producing AIC events 
in the $\log h-\log f_{\rm GW}$ plane,  in which
the GW frequency of double WDs $f_{\rm GW}=2/P_{\rm orb}$ 
where $P_{\rm orb}$ is the orbital period of the double WDs at the moment of their formation.
As previous studies in Ruiter et al. (2019), we simply assume that the distance $d$ from double WDs to the detectors is 
8.5\,kpc that is an appropriate distance from our solar system to the Galactic center.
In reality, the distance from the double WDs to the detectors  should have a 
wider distribution  in the Galaxy but not simply set to be 8.5\,kpc.

The GW strain amplitude $h$ in Fig.\,14 ranges from  $10^{\rm -24}-10^{\rm -21}$ 
and $f_{\rm GW}$ is in the range of $10^{\rm -5}-10^{\rm -1}$\,Hz.
For different kinds of double WDs,
there are two parts for the dimensionless $h$ distributions, 
in which the right small part  originates from Scenario C and the left part 
mainly originates from the Scenarios A and B in Figs\,9$-$10 (for the double CO WDs, see Fig. 11 of Wang 2018b). 
Furthermore, the $h$ and  $f_{\rm GW}$ for double ONe WDs is larger than that of ONe WD$+$CO WD systems, and 
 the $h$ and  $f_{\rm GW}$ for ONe WD$+$CO WD systems tend to be larger than that of double CO WDs.
We estimate that more than half of double WDs for producing AIC events are capable to be observed 
by the future space-based GW detectors, such as LISA, TianQin and Taiji, etc.

\section{Detection of AIC events}

Identification of AIC events is challenging. 
Up to now,  there is still no exclusively direct  evidence for AIC events in the observations. 
Some possible reasons are summarized as below:
(1)
AIC events are expected to be  relatively faint optical transients, which are fainter than a typical normal SNe Ia  by 5 mag or more
and last for only a few days to a week, indicating that this kind of objects are difficult to be discovered. 
(2)
AIC events with low luminosities may have been already observed by some ongoing surveys, 
but they were  mixed in some other (optically) faint transients, such as  
rapidly declining SN-like transients, fast-rising blue optical transients, kilonovae, 
fast radio bursts and gamma-ray bursts, etc. 
(3)
The predicted ejecta mass  and the $^{56}{\rm Ni}$ synthesized during the collapse 
are still not well determined. 
Another weakness for the AIC scenario is that it is difficult to reproduce the surface B-field and the spin rate 
of the  observed NSs that may be associated with AIC (see Tauris  et al. 2013).
The AIC scenario results in the production of normal NSs that then continue to accrete material, 
leading to a weaker B-field and a faster spin.
This is contrary with the observed NSs (characterized by  relatively high B-fields and slow spin) 
that are potential candidates for having formed through AIC in globular clusters or in the Galactic disk
(see Table 1 of Tauris  et al. 2013).

For the single-degenerate model, a dense circumstellar matter may be formed around the pre-AIC systems.
The ejecta from the AIC collides with the dense circumstellar matter, probably forming a strong shock.
The strong shock can produce synchrotron emission that may be detected in radio frequencies.
Piro \& Thompson (2014)  suggested that a strong radio emission would happen for an AIC event 
that originates from the evolution of ONe WD+RG systems. They
claimed that the ejecta of the AIC will collide with the surface of the RG star, producing a  strong X-ray emission lasting for $\sim$1\,hr 
followed by an optical signal from the shocked region, although the signal is strongly dependent on the viewing angle. 
Moriya (2016) suggested that AIC events may act as radio-bright but optically faint transients,  
likely leading to strong X-ray emission owing to
the interaction between AIC ejecta and  dense circumstellar matter in the single-degenerate 
systems. The radio signals of AIC are expected to be detected by the radio transient surveys such as 
the Very Large Array Sky Survey and the Square Kilometer Array transient survey (e.g., Moriya 2016).

For the double-degenerate model,  it is not expected to have dense circumstellar matter before AIC, 
thus no strong radio emission as predicted in the single-degenerate model.
As shown in Fig.\,12,  the merging of double WDs can form super-massive ($>$$2.2\,M_{\odot}$) NSs 
that require rapid rotation to support themselves (e.g., Metzger et al. 2015). 
The super-massive NSs may lose their rotational energy immediately 
through the r-mode instability and collapse into black holes finally.
It has been suggested that the collapsing of super-massive NSs  may produce 
fast radio bursts if they are strongly magnetized (e.g., Moriya 2016).
Lyutikov \& Toonen (2017) proposed an alternative way to form  prompt 
short gamma-ray bursts  following the merging of ONe WD+CO WD.
Rueda et al. (2019) predicted that the primary WD during the 
the post-merger time may appear as a pulsar, 
depending on the rotation period and the value of the magnetic fields.
In addition, Yu et al. (2019a) recently suggested that searching for 
dust-affected optical transients and shock-driven radio transients 
will help to unveil the nature and evolution of  double WD mergers with super-${M}_{\rm Ch}$.

Anderson et al. (2019)  reported  a radio transient  VTC J095517.5$+$690813  that
became bright in the radio but without an optical counterpart in the observations.
By comparing the predicted radio emission from an AIC with that from VTC J095517.5$+$690813,  
Moriya (2019) recently suggested that the radio transient can be reasonably explained by AIC of a WD.
In addition,  Scholz et al. (2016)  discovered a persistent radio source associated with fast radio burst FRB 121102. 
The persistent radio source has been thought to be produced by a weak stellar 
explosion with a small ejected mass that may be consistent with the AIC scenario, 
whereas the resulting NS acts as the energy source of FRB 121102 (e.g., Waxman 2017; Sharon \& Kushnir 2020).
Margalit et al. (2019) recently also associated the progenitor of FRB 121102 with AIC. 
Therefore,  there is a possibility that AIC events might be first discovered as 
fast radio bursts, and then followed by optically faint but radio-bright transients (e.g., Moriya 2019).
Additionally, Moriya (2019) recently speculated that 
it is possible to distinguish the single-degenerate model and the double-degenerate 
models via the radio light curves once an AIC event has been identified.

AT2018cow was discovered by the Asteroid Terrestrial-impact Last Alert System (ATLAS; see Tonry et al. 2018).
It is likely to be the brightest member of the class of fast-rising blue optical transients (e.g., Prentice et al. 2018; Margutti et al. 2019).
Yu et al. (2015) associated the AIC scenario with the recently discovered fast-evolving luminous transients. In a further work, 
Yu et al. (2019b)  predicted hard and soft X-ray emissions from the AIC of WDs, 
providing a clear observational feature for identifying AIC events in future observations, especially for AT2018cow.
Lyutikov \& Toonen (2019) recently argued that the fast-rising blue optical transients and its brightest member AT2018cow may result from  
the merging of an ONe WD with another WD. Note that 
Soker et al. (2019) proposed a common envelope jets SN scenario for the formation of AT2018cow, in which
a jet driven by an accreting NS collides with a giant star. 
Note also that  the unusual transient AT2018cow was claimed as a WD tidal disruption event (see Kuin et al. 2019). 

Furthermore, the faint rapidly declining SN-like transients are possible candidates of AIC events.
Apart from  the most detailed observed kilonova AT2017gfo, McBrien et al. (2019) recently suggested that
SN 2018kzr is the second fastest declining SN-like transient known so far, which was independently 
discovered  by the Zwicky Transient Facility (ZTF; see Bellm et al. 2019) and  the ATLAS survey (see Tonry et al. 2018). 
By modelling the bolometric light curves and early spectra, McBrien et al. (2019) estimated that 
SN 2018kzr has a low ejecta mass composed of intermediate mass elements, disfavouring the merging of double NSs. 
Thus, they argued that AIC of an ONe WD may  provide an alternative explanation for the formation of 
the rapidly declining transient. 

\section{Summary}

The AIC scenario has been proposed as 
a theoretically predicted outcome of ONe WDs when they grow in mass close to ${M}_{\rm Ch}$,
relating to the formation of NS systems.  
So far, there has been no direct detection of any AIC event, 
likely because this kind of events are relatively faint optical transients and short-lived.
Currently, the most studied evolutionary pathways for AIC events 
are the single-degenerate model  and the double-degenerate model. 
We review recent progress on the two classic progenitor models of AIC events.
We also  review recent results of binary population synthesis and summarize
recent theoretical and observational constraints on the detection of AIC events.

In the single-degenerate model, the pre-AIC systems could potentially be identified 
as supersoft X-ray sources, symbiotics and 
cataclysmic variables.
The post-AIC systems in the  single-degenerate model may evolve to some different kinds NS systems if AIC happens, 
mainly depending on the chemical components of the mass donors, as follows:
(1)  The post-AIC systems with MS donors is likely to be identified as LMXBs and 
the resulting fully recycled MSPs with He WD companions.
(2) The post-AIC systems with RG donors could form LMXBs and the more mildly recycled MSPs 
with He/CO WD companions, especially for MSPs with much longer orbital periods compared with MS donors.
(3) The post-AIC systems with He star donors could potentially be identified as 
IMXBs and eventually produce IMBPs with short orbital periods. 

In the double-degenerate model, the pre-AIC systems are close double WDs with short orbital periods. 
The double WDs for producing AIC events are capable to be observed by
the future space-based GW detectors, such as LISA, TianQin and Taiji, etc.
The post-AIC systems in the classic double-degenerate model
are single isolated NSs  that may correspond to a specific kind of NSs and 
show some peculiar properties compared with those from core-collapse SNe. 

AIC events are likely to be radio-bright but optically faint.
In order to provide constraints on the progenitor models of AIC events, 
large samples of WD binaries are required  in the observations.
Meanwhile, 
identifying AIC events in future transient surveys and
understanding their electromagnetic signals are important for the confirmation of the AIC process.
More theoretical and observational investigations would be helpful for our understanding of this type of events.
Future observational surveys are expected to finally confirm such events with low luminosities 
and to clarify the long-term issue that the current stellar evolution theories confront.

\begin{acknowledgements}
We acknowledge the referee for the valuable comments that helped us to improve this review. 
We also thank Noam Soker, Wencong Chen, Takashi Moriya, Xiangcun Meng, Matthias U. Kruckow, Iminhaji Ablimit, 
Zongkuan Guo, Hailiang Chen, Jujia Zhang,  Yiming Hu, Haijun Tian, Yan Gao and Wenshi Tang
for their helpful discussions and suggestions.
BW is supported by the National Natural Science Foundation of China (Nos 11521303, 11673059 and 11873085),
the Chinese Academy of Sciences (No QYZDB-SSW-SYS001),
and the Yunnan Province (Nos 2018FB005 and 2019FJ001). 
DL is supported by the Natural Science Foundation of China (No 11903075) and 
the Western Light Youth Project of Chinese Academy of Sciences.

\end{acknowledgements}

\label{lastpage}
\end{document}